\documentclass{aa}  

\usepackage{graphicx}
\usepackage{txfonts}
\usepackage{lipsum}
\usepackage{subcaption}  
\usepackage{lscape}  
\usepackage{placeins}    

\begin{document}

   \title{Characterizing the physical and chemical properties of the Class I protostellar system Oph-IRS 44}

   \subtitle{Binarity, infalling streamers, and accretion shocks}

   \author{E. Artur de la Villarmois \inst{1} 
      \and V. V. Guzm\'an \inst{2,3} 
      \and M. L. van Gelder \inst{4} 
      \and E. F. van Dishoeck \inst{4}      
      \and E. A. Bergin \inst{5} 
      \and D. Harsono \inst{6} 
      \and N. Sakai \inst{7} 
      \and J. K. J{\o}rgensen \inst{8} 
        }

   \institute{European Southern Observatory, Alonso de C\'ordova 3107, Casilla 19, Vitacura, Santiago, Chile\\
      \email{Elizabeth.ArturdelaVillarmois@eso.org}
      \and Instituto de Astrof\'isica, Pontificia Universidad Cat\'olica de Chile, Av. Vicu\~{n}a Mackenna 4860, 7820436 Macul, Santiago, Chile
      \and N\'ucleo Milenio de Formaci\'on Planetaria (NPF), Chile
      \and Leiden Observatory, Leiden University, P.O. Box 9513, 2300RA Leiden, The Netherlands
      \and Department of Astronomy, University of Michigan, 1085 S. University Ave., Ann Arbor, MI 48109, USA 
      \and Institute of Astronomy, Department of Physics, National Tsing Hua University, Hsinchu, Taiwan
      \and RIKEN Cluster for Pioneering Research, 2-1, Hirosawa, Wako-shi, Saitama 351-0198, Japan
      \and Niels Bohr Institute, University of Copenhagen, {\O}ster Voldgade 5-7, 1350, Copenhagen, Denmark\\ }

   \date{Received August 21, 2025}

   \abstract
   {In the low-mass star formation process, theoretical models predict that material from the infalling envelope could be shocked as it encounters the outer regions of the disk. This is followed by an increase in the dust temperature and sublimation, into the gas phase, of molecular species that will otherwise remain locked on dust grains. Although accretion shocks are predicted by theoretical models, only a few protostars show evidence of these shocks at the disk-envelope interface, and the main formation path of shocked-related species is still unclear. They can be formed entirely on dust surfaces and then sublimated, or through reactions in the gas phase, or a combination of both.}   
   {The goal of this work is to assess the chemistry associated with accretion shocks and the formation path of molecules that are usually associated with these dense and warm regions.}
   {We present new observations of IRS 44, a Class I source with a resolved disk that has previously been associated with accretion shocks, taken at high angular resolution (0$\farcs$1, corresponding to 14~au) with the Atacama Large Millimeter/submillimeter Array (ALMA). We observe three different spectral settings in bands 6 and 7, targeting multiple molecular transitions of CO, H$_{2}$CO, and simple sulfur-bearing species (such as CS, SO, SO$_{2}$, H$_{2}$S, OCS, and H$_{2}$CS).}
   {In continuum emission, the binary nature of IRS 44 is observed for the first time at sub-millimeter wavelengths and the emission agrees with the optical and infrared counterparts. Infalling signatures are seen for the CO 2--1 line and the emission peaks at the edges of the continuum emission around IRS 44 B, the same region where bright SO and SO$_{2}$ emission is seen. Weak CS and H$_{2}$CO emission is observed, while OCS, H$_{2}$S, and H$_{2}$CS transitions are not detected. }
   {IRS 44 B seems to be more embedded than IRS 44 A, indicating a non-coeval formation scenario or the rejuvenation of source B due to late infall. CO 2--1 emission is tracing the outflow component at large scales, infalling envelope material at intermediate scales, and two infalling streamer candidates are identified at disk scales. Infalling streamers might produce accretion shocks when they encounter the outer regions of the infalling-rotating envelope. These shocks heat the dust and efficiently release S-bearing species (such as H$_{2}$S, SO, and SO$_{2}$), as well as promoting a lukewarm chemistry ($\sim$200~K) in the gas phase. With the majority of carbon locked in CO, there is little free C available to form CS and H$_{2}$CS in the gas, leaving an oxygen-rich environment. The high column densities of SO and SO$_{2}$ might therefore be a consequence of two processes: direct thermal desorption from dust grains and gas-phase formation due to the availability of O and S. IRS 44 is an ideal candidate with which to study the chemical consequences of accretion shocks and the dynamical connections between the envelope and the disk, through infalling streamers.  }
   

   \titlerunning{Characterizing the physical and chemical properties of Oph-IRS 44.}
   \maketitle

\section{Introduction}

The formation and evolution of low-mass protostars and their disks are fundamental to understanding the formation of our own Solar System. A typical low-mass star forms when a molecular cloud with angular momentum collapses, the central protostar accretes mass, material from the envelope infalls into the central regions, and a circumstellar disk forms in the equatorial plane \citep{Cassen1981, Terebey1984, Shu1993, Hartmann1998}. Eventually, planets will form within the disk, and their final composition is strongly dependent on the physical and chemical processes within the circumstellar disk \citep[e.g.,][]{Herbst2009, Drozdovskaya2018, Jorgensen2020, Oberg2023, vantHoff2024}. In recent years, it has been found that infalling and accretion processes are not isotropic. How protostars accrete material from the surrounding envelope is still an open question \citep[e.g., ][]{Pineda2023, Kuffmeier2023, Kuffmeier2020}. 

Theoretical models predict that streamer-like infall supply material from the envelope (or molecular cloud scales) onto the disk, and accretion shocks are created at the envelope-disk interface \citep{Ulrich1976, Mendoza2009, Kuffmeier2019}. These shocks induce an increase in the temperature and species that formed in grain mantles are subsequently released into the gas phase, affecting the chemical evolution of the early disk and the material available for planet formation \citep{vanGelder2021}. The chemical content of the early disk could therefore be partially reset after the passage of the shock, whereas the absence of shocks suggests a chemical inheritance between the envelope and the disk. Despite being a natural consequence in theoretical models, only a few low-mass protostars show observational evidence of streamers and accretion shocks \citep[e.g.,][]{Sakai2014, Yen2019, Pineda2020, Artur2022, Garufi2022, Valdivia2022, Hsieh2023, Gupta2024, Liu2025}. Furthermore, it is still not well understood if the observed shock-related species are being formed entirely on the dust surfaces and then sublimated, or if there is an important contribution to the formation of the molecule in the gas phase \citep{vanGelder2021}.

IRS 44, also known as YLW 16 \citep{Young1986}, is a Class I source located in the Ophiuchus molecular cloud, at a distance of 139 pc \citep[average value for the L1688 cloud;][]{Canovas2019}. It has been proposed that IRS 44 is a protobinary system with a separation of 0$\farcs$3, based on optical and infrared observations \citep{Allen2002, Duchene2007, Herczeg2011}. Nevertheless, \cite{Sadavoy2019}, \cite{Artur2019a}, and \cite{Artur2022} did not find any evidence of binarity in ALMA data at an angular resolution of 0$\farcs$25 (35~au) in band 6, 0$\farcs$4 (56~au) in band 7, and 0$\farcs$1 (14~au) in band 7, respectively. Strong and compact SO$_{2}$ emission was first detected toward IRS 44 by \cite{Artur2019a} and, later on, IRS~44 was associated with accretion shocks through the detection of multiple SO$_{2}$ transitions \citep{Artur2022}. The latter work provided the following physical properties for the inner regions of IRS 44 ($\leq$50~au): H$_{2}$ densities higher than 10$^{8}$~cm$^{-3}$, SO$_{2}$ rotational temperatures between 90 and 250~K, and high SO$_{2}$ column densities of between 0.4 and 1.8~$\times$~10$^{17}$~cm$^{-2}$. IRS 44 is therefore a suitable source in which to search for other sulfur-bearing species, assess the main formation path of SO$_{2}$, and understand the chemistry related to accretion shocks and potential infalling streamers.

In this paper we present high-angular-resolution 0$\farcs$1 (14~au) ALMA observations of multiple molecular transitions toward IRS 44, most of them related with sulfur-bearing species. Section 2 describes the observational procedure, calibration, and CO 2--1 archival data. The observational results are presented in Sect. 3, together with the observed molecular transitions and the interpretation of a streamer candidate. Section 4 is dedicated to the analysis of the data, with the estimation of temperatures and column densities. We discuss the chemistry related to IRS 44 in Sect. 5, and end with a summary in Sect. 6.

\section{Observations}

IRS 44 was observed with ALMA during April, May, and June 2023 as part of the program 2022.1.00209.S (PI: Elizabeth Artur de la Villarmois). At the time of the observations, between 30 and 44 antennas were available in the array providing baselines between 15 and 3697 m. The observations targeted three different spectral settings -- two of them in band 7 and one in band 6 -- to observe multiple transitions of sulfur-bearing species, CO isotopologs, and H$_{2}$CO. These lines are presented and discussed in Sect. 3.2.

The calibration and imaging were done in CASA\footnote{http://casa.nrao.edu/} pipeline version 6.4.1 \citep{McMullin2007}. Gain and bandpass calibrations were performed through the observation of the quasars J1554--2704, J1617--2537, and J1700--2610. Imaging was performed using the \texttt{tclean} task in CASA, where the Briggs weighting with a robust parameter of 0.5 was employed. The automasking option was chosen and the velocity resolution is 0.2~km~s$^{-1}$. Self-calibration was performed only for the continuum data, where the final solution intervals, central frequencies, beam size, and root mean square (rms) values are listed in Table~\ref{table:calibration} for the three spectral settings. The largest angular scale (LAS) of these observations is $\sim$1$\farcs$2 for both bands, band 6 and band 7, as band 6 was observed in configuration C43-7 and band 7 in C43-6.

\subsection*{Archival ALMA data}
Only one data cube was retrieved from the ALMA archive, which corresponds to the CO 2--1 transition that is part of project 2019.1.01792.S (PI: Diego Mardones). We took the product cube from the archive, without additional calibration or cleaning. The synthesized beam is 1$\farcs$2~$\times$~0$\farcs$9, the LAS is 5$\farcs$0, and the rms is 0.016~Jy~beam$^{-1}$. The main purpose of retrieving this cube was to assess the CO intermediate spatial scales (between 1$\farcs$0 and 5$\farcs$0) that are filtered out in our data (LAS = 1$\farcs$0), but this dataset was not used in the analysis. The red- and blueshifted contours for the archival data are presented in Appendix~\ref{Ap1}.

\begin{table*}[h!]
        \caption{Parameters of the continuum observations after applying self-calibration.}
        \label{table:calibration}
        \centering
        \begin{tabular}{c c c c c}
                \hline\hline
                	Central frequency	& Final sol\_int (self-cal)		& Beam size			& Beam PA*	& rms				\\  
                	(GHz)                     	& (s)              				& ("~$\times$~")		& ($\degr$)	& (mJy~beam$^{-1}$)	\\
                \hline		
		233.0			& 60.48					& 0.11~$\times$~0.09	& 91.2		& 0.01				\\
		303.0			& 30.24					& 0.11~$\times$~0.10	& 78.3		& 0.11				\\	
		328.2			& 30.24					& 0.13~$\times$~0.12	& 98.5		& 0.22				\\	
                \hline
        \end{tabular}
        \tablefoot{*Position angle (PA) is measured from north to east.}
\end{table*}

\begin{figure*}[h!]
        \centering
        \includegraphics[width=.98\textwidth]{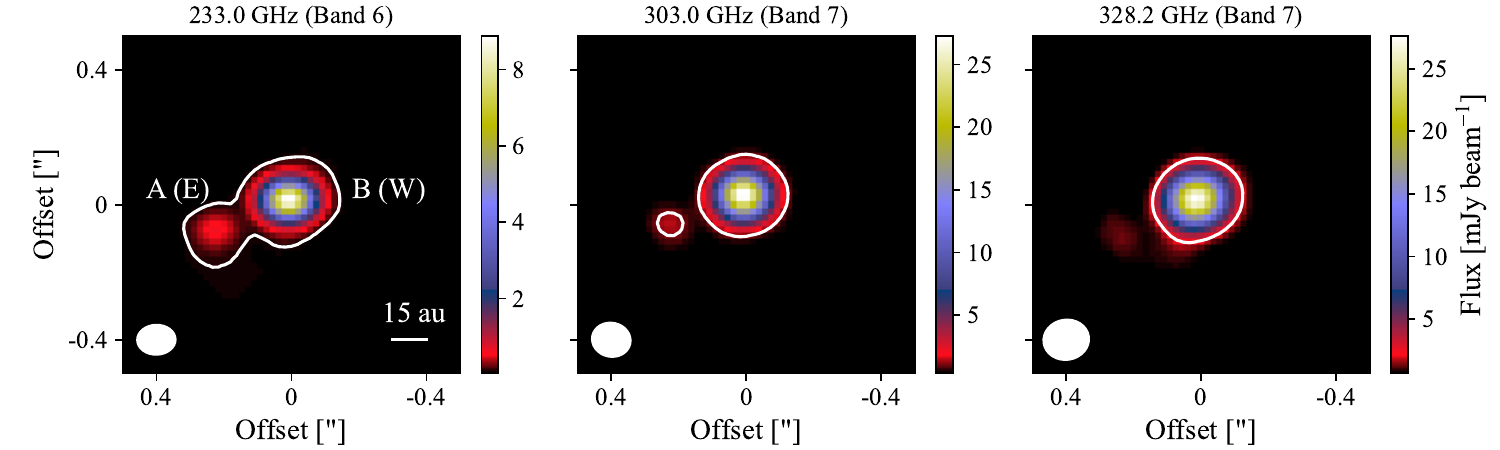}
        \caption[]{\label{fig:cont}
        Continuum emission of IRS 44 above 3$\sigma$ at different frequencies. The white contour represents a flux value of 10$\sigma$. The synthesized beam is represented by a filled white ellipse in the lower left corner of each panel. IRS~44 A and B can also be found as IRS~44 E and W in the literature, respectively \citep[e.g.,][]{Herczeg2011}.
        }
\end{figure*}

\begin{figure}[h!]
        \centering
        \includegraphics[width=.42\textwidth]{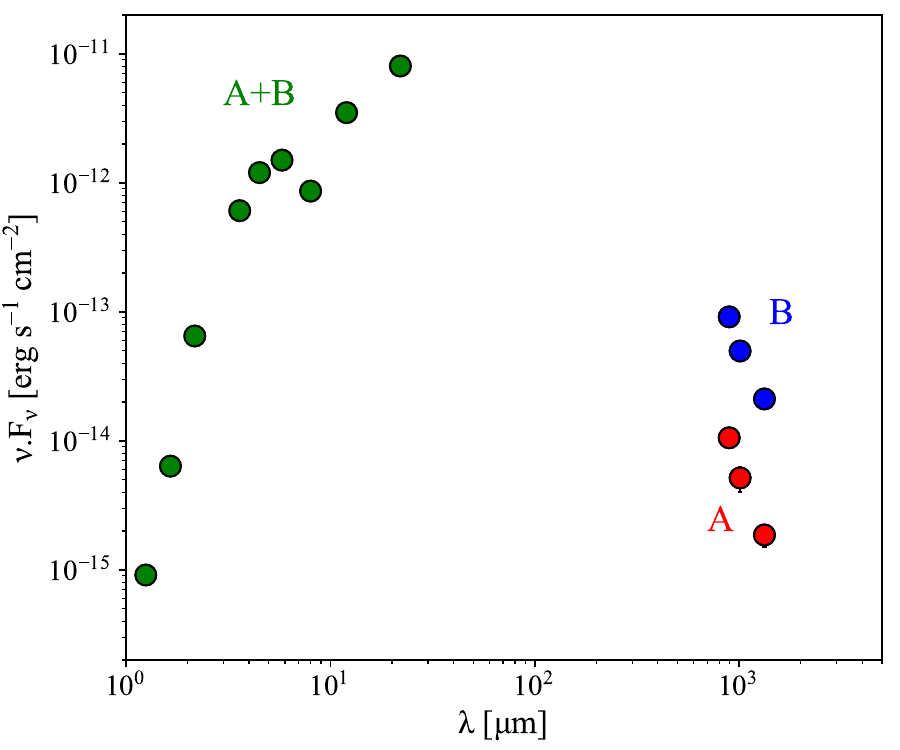}
        \caption[]{\label{fig:sed}
        Spectral energy distribution (SED) of IRS 44. In the infrared regime (green dots), the flux corresponds to both sources, source A being the brightest one \citep{Terebey2001,Dunham2015}. In ALMA bands, the binary system is resolved and source B (blue dots) is $\sim$10 times brighter than source A (red dots). 
        }
\end{figure}

\begin{table}[h!]
        \caption{Integrated and peak continuum fluxes for both sources.}
        \label{table:cont}
        \centering
        \begin{tabular}{c c c}
                \hline\hline
                	Frequency		&Flux$_\mathrm{integrated}$		& Flux$_\mathrm{peak}$	\\  
                	(GHz)                     	& (mJy)               				& (mJy/beam)	         	\\
                \hline		
                \multicolumn{3}{c}{\textbf{Source A}}										\\
                \hline									
		233.0			& 0.83~$\pm$~0.17				& 0.60~$\pm$~0.01		\\
		303.0			& 1.74~$\pm$~0.39				& 1.50~$\pm$~0.11		\\	
		328.2			& 3.15~$\pm$~0.44				& 1.30~$\pm$~0.22		 \\	
                \hline		
                \multicolumn{3}{c}{\textbf{Source B}}										\\
                \hline											
		233.0			& 9.38~$\pm$~0.10				& 8.97~$\pm$~0.01 		\\
		303.0			& 16.76~$\pm$~0.26				& 14.70~$\pm$~0.11		\\	
		328.2			& 27.28~$\pm$~0.63				& 23.24~$\pm$~0.22 	\\	
		
                \hline
        \end{tabular}
\end{table}

\begin{figure*}[htbp]
        \centering
        \includegraphics[width=.95\textwidth]{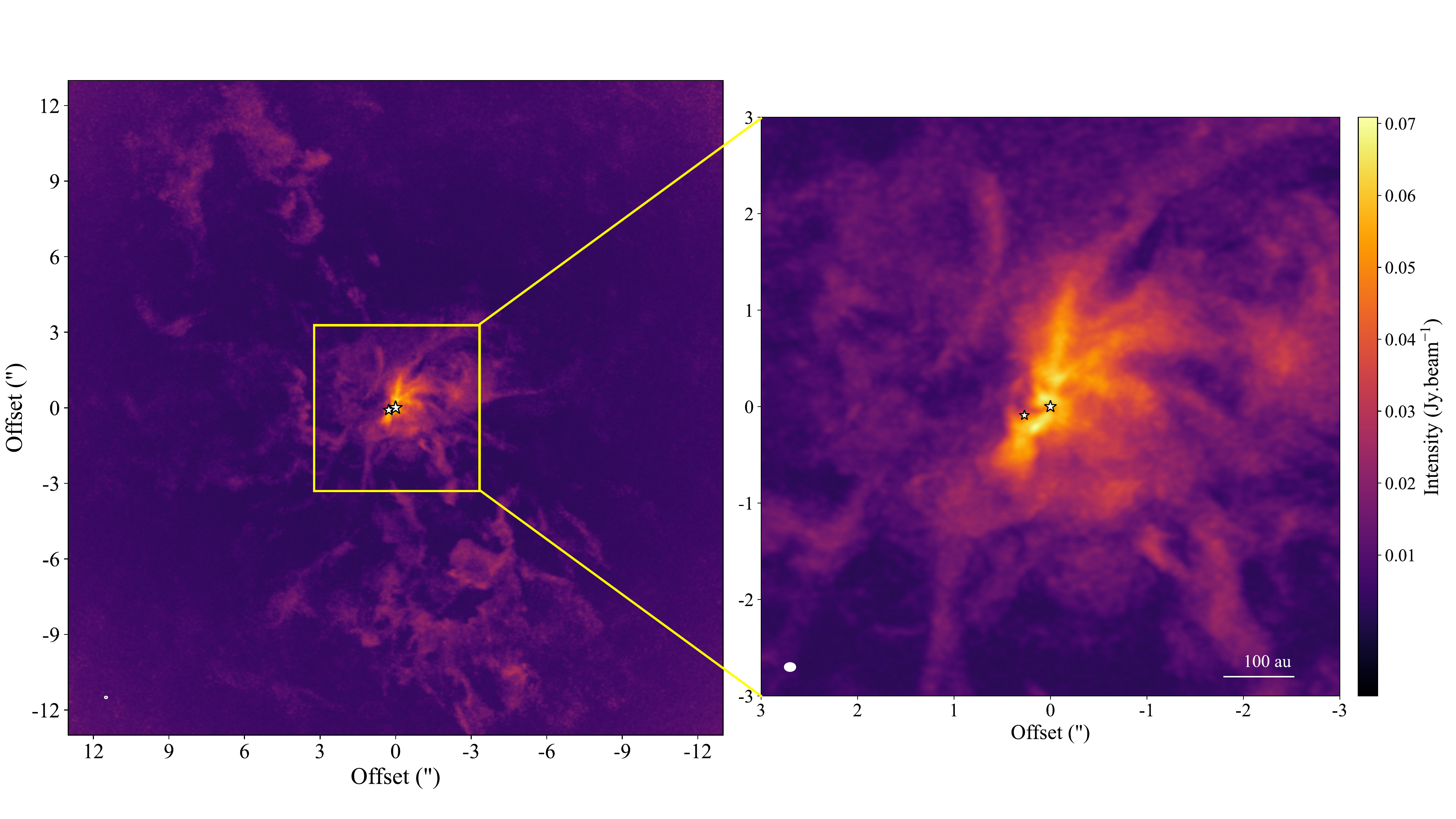}
        \caption[]{
        CO 2--1 emission (moment 8) at large (\textit{left}) and intermediate scales (\textit{right}). The white stars represent the position of the A and B components (see Fig.~\ref{fig:cont}). The synthesized beam is represented by a filled white ellipse in the lower left corner of each panel.  
        }
        \label{fig:CO_mom8}
\end{figure*}

\begin{figure*}[h!]
        \centering
        \includegraphics[width=.98\textwidth]{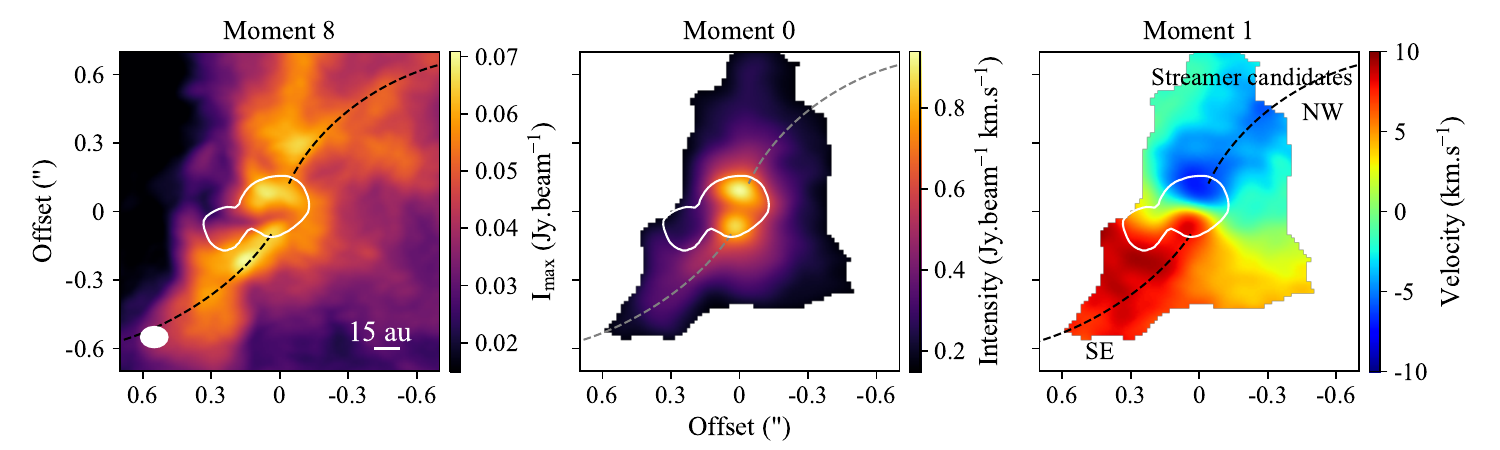}
        \caption[]{\label{fig:CO_moms}
        CO 2--1 emission at small scales. \textit{Left}: Maximum value map (moment 8) above a 3$\sigma$ level. \textit{Center}: Integrated map (moment 0) above a 15$\sigma$ level (1$\sigma$~=~10~mJy~beam$^{-1}$~km~s$^{-1}$). \textit{Right}: Velocity map (moment 1) above a 15$\sigma$ level. The white contour represents the continuum emission at 233~GHz at a 10$\sigma$ value and the dashed black curves indicate the direction of the proposed streamers. The synthesized beam is represented by a filled white ellipse in the left panel.    
        }
\end{figure*}

\begin{figure*}[h!]
        \centering
        \includegraphics[width=.98\textwidth]{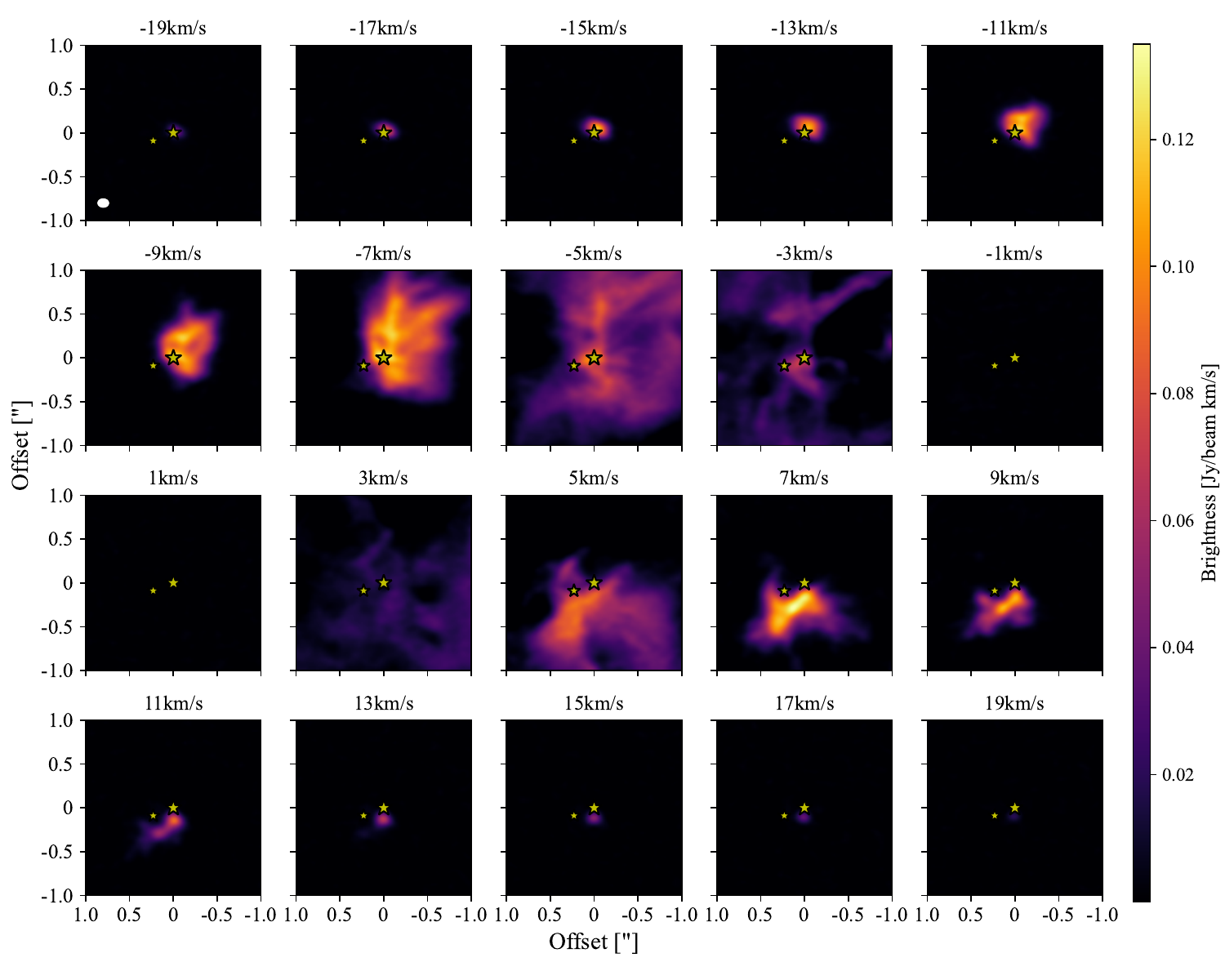}
        \caption[]{\label{fig:CO_chans}
        Velocity channel maps for CO 2--1 above 1$\sigma$. The systemic velocity (3.7~km~s$^{-1}$) is shifted to zero and each map has a velocity width of 2~km~s$^{-1}$. The yellow stars show the position of IRS44~A and IRS44~B, while the synthesized beam is represented by a filled white ellipse in the upper left panel.
        }
\end{figure*}

\begin{figure}[h!]
        \centering
        \includegraphics[width=.49\textwidth]{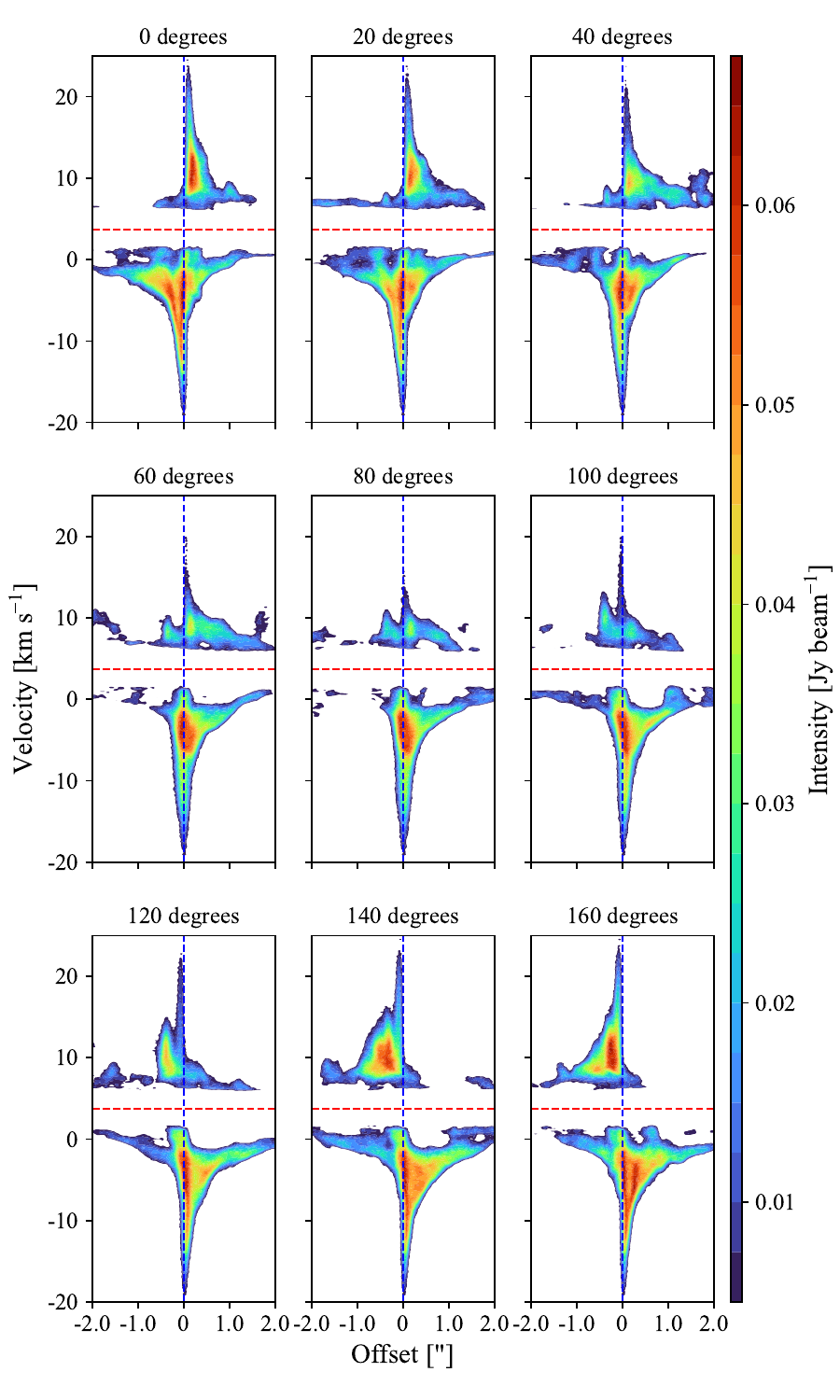}
        \includegraphics[width=.4\textwidth]{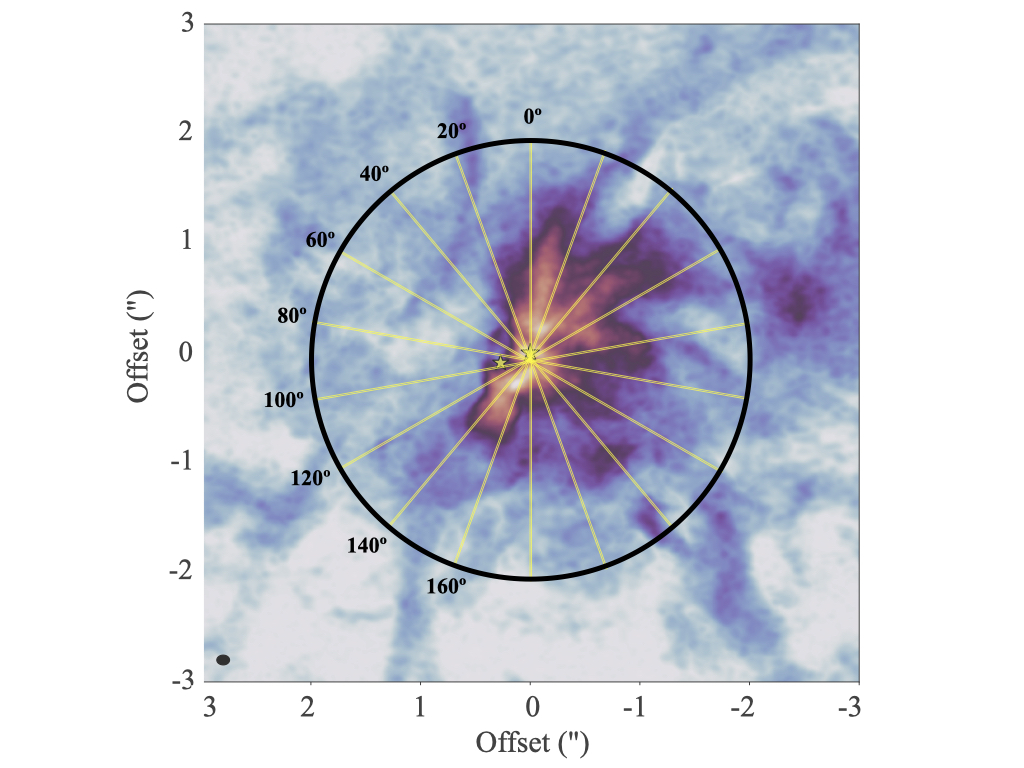}
        \caption[]{\label{fig:CO_PV}
        Position-velocity (PV) diagrams for CO 2--1 (\textit{upper panels}) above 3$\sigma$. The horizontal dashed red line represents the systemic velocity of 3.7~km~s$^{-1}$, while the vertical dashed blue line corresponds to the central position of source B. The different PAs (\textit{lower panel}) are indicated by yellow lines over the CO 2--1 moment 8 map.  
        }
\end{figure}

\section{Results}
\subsection{Continuum emission}
\subsubsection*{IRS~44: A proto-binary}

The continuum emission is shown in Fig.~\ref{fig:cont}, for three different frequencies (233.0, 303.0, and 328.2~GHz). Source A is clearly detected in ALMA band 6, with a signal-to-noise ratio (S/N) of around 60, while the S/N decreases for shorter wavelengths (S/N~=~14 and 6 for 303.0 and 328.2~GHz, respectively). Although the peak intensity increases for band 7 observations, the rms does as well, decreasing the S/N (see Table~\ref{table:cont}). The integrated and peak flux were calculated using two-dimensional (2D) Gaussian fits in the image plane. Furthermore, the protostellar disk mass, \textit{M$_\mathrm{disk}$}, was calculated for both sources, using the following equation:

\begin{equation} 
    M_\mathrm{disk} = \frac{S_{\nu}d^{2}}{\kappa_{\nu}B_{\nu}(T)} \ ,
    \label{eq:Eq2}
\end{equation}

\noindent where \textit{S$_{\nu}$} represents the surface brightness, \textit{d} the distance to the source, $\kappa$$_\mathrm{\nu}$ the dust opacity, and B$_\mathrm{\nu}$(\textit{T}) the Planck function for a single temperature. The adopted dust temperature (\textit{T$_\mathrm{dust}$}) is 15~K \citep[see][]{Dunham2014b}, and the derived masses (gas~+~dust, assuming a gas-to-dust ratio of 100) are listed in Table~\ref{table:cont_mass}. We find that the disk around source B is 10 times more massive than that of source A. Additionally, there is a difference between the disk mass around source B for different wavelengths, where the highest value corresponds to the longer wavelength. This discrepancy hints at optically thick emission, mainly at 303.0 and 328.2~GHz, but the continuum emission at 233.0~GHz might also suffer from optical depth effects (to a lesser degree). Therefore, \textit{M$_\mathrm{disk}$} should be taken as a lower limit.

\begin{table}[h!]
        \caption{Calculated disk masses for both sources, A and B, from the continuum fluxes.}
        \label{table:cont_mass}
        \centering
        \begin{tabular}{c c c c c}
                \hline\hline
                	Frequency		& $\lambda$	& $\kappa_{\nu}$ $^{(1)}$		& \textit{M$_\mathrm{B}$} $^{(2)}$			& \textit{M$_\mathrm{A}$} $^{(2)}$		\\
                	(GHz)                     	& (mm)		& (cm$^{2}$.g$^{-1}$)	        	& ($\times$~10$^{-3}$~M$_{\odot}$)	& 		$ (\times$~10$^{-3}$~M$_{\odot}$)		\\
                \hline		
		233.0			& 1.287		& 0.0092					& 5.6~$\pm$~0.2						& 0.5~$\pm$~0.3				\\
		303.0			& 0.989		& 0.0141					& 4.4~$\pm$~0.2						& 0.5~$\pm$~0.3				\\	
		328.2			& 0.913		& 0.0172					& 5.2~$\pm$~0.4						& 0.6~$\pm$~0.3				\\	
                \hline
        \end{tabular}
        \tablefoot{$^{(1)}$ From \cite{Ossenkopf1994} using OH5. $^{(2)}$ Total mass (gas + dust).}
\end{table}

Figure \ref{fig:sed} shows the spectral energy distribution (SED) of IRS 44 as a system. Source A is brighter in the infrared regime \citep{Terebey2001,Dunham2015}, while the opposite situation is seen at sub-millimeter wavelengths where both components are resolved: source B is much brighter than source A. This suggests that \textit{i)} both sources are in different evolutionary stages, \textit{ii)} they have different orientations, or \textit{iii)} source B appears younger due to late infall \citep[e.g.,][]{Kuffmeier2023}. \cite{Murillo2016} found that 33$\%$ of multiple protostellar systems are non-coeval, mainly due to its formation history and different dynamical evolution. Given that a dark lane is observed along source B in optical images \citep[see Fig. 2 in][]{Terebey2001} and the fact that molecular transitions, such as CO, are not detected toward source A \citep[e.g.,][]{Herczeg2011}, a non-coeval scenario or a rejuvenation of source B due to late infall are the more plausible scenarios for the IRS~44 system. The binarity of IRS~44 is better seen in band 6 observations, which have been targeted before by \cite{Sadavoy2019} but with a poorer angular resolution (0$\farcs$25) that is not enough to resolve both sources. On the other hand, \cite{Artur2022} observed IRS~44 in band 7 (330~GHz) with the same angular resolution of 0$\farcs$1; however, the detection of source A was not clear, similar to in the right panel of Fig.~\ref{fig:cont}.

\subsection{Molecular transitions}

The ALMA observations consist of three spectral settings targeting several molecular transitions, which are listed in Table~\ref{table:molecules}. CO isotopologs, SO, $^{34}$SO, and SO$_{2}$, are among the detected species, together with CS and only one clear detection of H$_{2}$CO. On the other hand, transitions from $^{34}$SO$_{2}$, OCS, H$_{2}$S, and H$_{2}$CS are not detected at a 3$\sigma$ level. 

Figure~\ref{fig:CO_mom8} shows the moment 8 map (maximum value of the spectrum) of the CO 2--1 transition at large and intermediate scales. The same map is superimposed with the CO 2--1 velocity contours from the archival data (see Figure~\ref{fig:CO_mom8_archive} in the appendix), which has a larger beam size (1$\farcs$2~$\times$~0$\farcs$9) and a LAS of 5$\farcs$0. The CO emission seems to trace arc-like structures at small scales, closer to the binary system, an elongated envelope structure encompassing the arcs, and an extended component in the northeast--southwest direction. The latter component is consistent with the outflow direction (PA~=~20$\degr$) observed by \cite{vanderMarel2013} using single-dish observations of CO 3--2. Given that the LAS of our observations related to the moment 8 map is 1$\farcs$0, emission more extended than that value will be filtered out by the interferometer; therefore, only the densest regions are seen.

\subsubsection{Streamer candidates} 

The arc-like structures could potentially be associated with infalling streamers but, given that significant emission is filtered out, a kinematic analysis is needed. Figure~\ref{fig:CO_moms} shows moment 8, 0, and 1 maps for the CO 2--1 transition in a zoomed-in-on area. The moment 8 presents three main arc-like structures in the northwest, while only one prominent component is seen toward the southeast. The moment 0 map shows more compact emission, covering an area of 0$\farcs$6, and a weaker tail is observed toward the southeast. On the other hand, the moment 1 map exhibits a clear velocity gradient toward source B, with two main components. The redshifted one coincides with the prominent component seen toward the southeast in the moment 8 map, which we call the SE streamer candidate. The blueshifted emission correlates with only one of the two arc-like structures seen in the northern regions of source B (see moment 8 map), which is the proposed NW streamer candidate. This velocity gradient is likely due to infalling gas, as higher velocities are seen closer to the protostar; this is discussed in more detail in the next few paragraphs. The black-dashed curve in Fig.~\ref{fig:CO_moms} represents the directions of the streamer candidates from this work, which are remarkably symmetric around the continuum peak of source B. A more detailed theoretical and kinematic analysis of the streamer candidates will be presented in a future work. The spectra and moment 0 maps of the CO isotopologs are shown in Fig.~\ref{fig:Spec_CO} in the appendix.

Figure~\ref{fig:CO_chans} shows channel maps for the CO 2--1 emission, for ranges of 2 km~s$^{-1}$. Maps with mean velocities between -7 and -11~km~s$^{-1}$ show that the blueshifted emission extends to $\sim$1$\farcs$0 and approaches source B as the velocity increases, in agreement with infalling signatures. Additionally, more compact emission is seen for higher velocities, between -13 and -19~km~s$^{-1}$. Redshifted emission shows a similar behavior to that of negative velocities, but the emission is more confined toward the southeast.

\begin{figure*}[h!]
	\sidecaption
        \includegraphics[width=12cm]{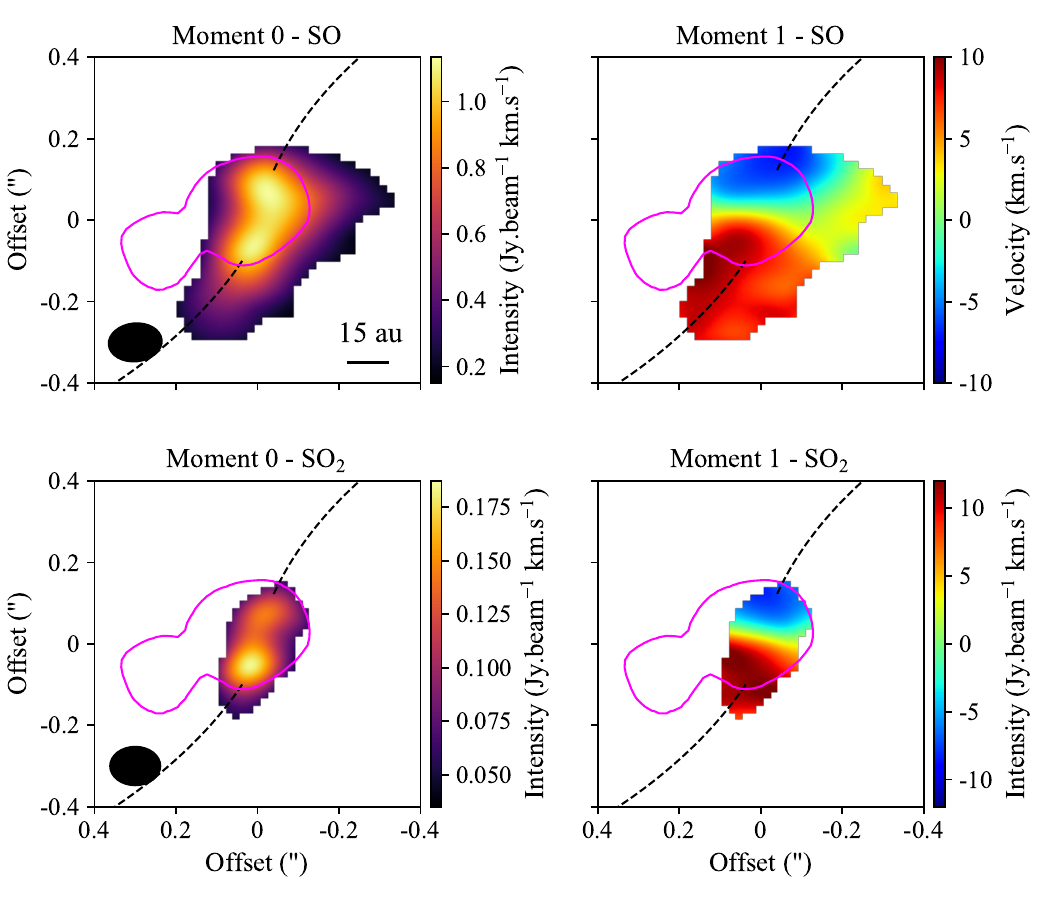}
        	\caption{
        SO and SO$_{2}$ emission. \textit{Upper panels}: Moment 0 and 1 maps for SO 7$_{8}$--6$_{7}$ above 5$\sigma$. \textit{Lower panels}: Moment 0 and 1 maps for SO$_{2}$ 22$_{2,20}$--22$_{1,21}$ above 5$\sigma$. The magenta contour represents the continuum emission at 233~GHz at a 10$\sigma$ value and the dashed black curves indicate the direction of the proposed streamers. The synthesized beam is represented by a filled black ellipse in the left panels.    
        }
        	\label{fig:SO_moms}
\end{figure*}

\begin{figure*}[h!]
	\sidecaption
        \includegraphics[width=12cm]{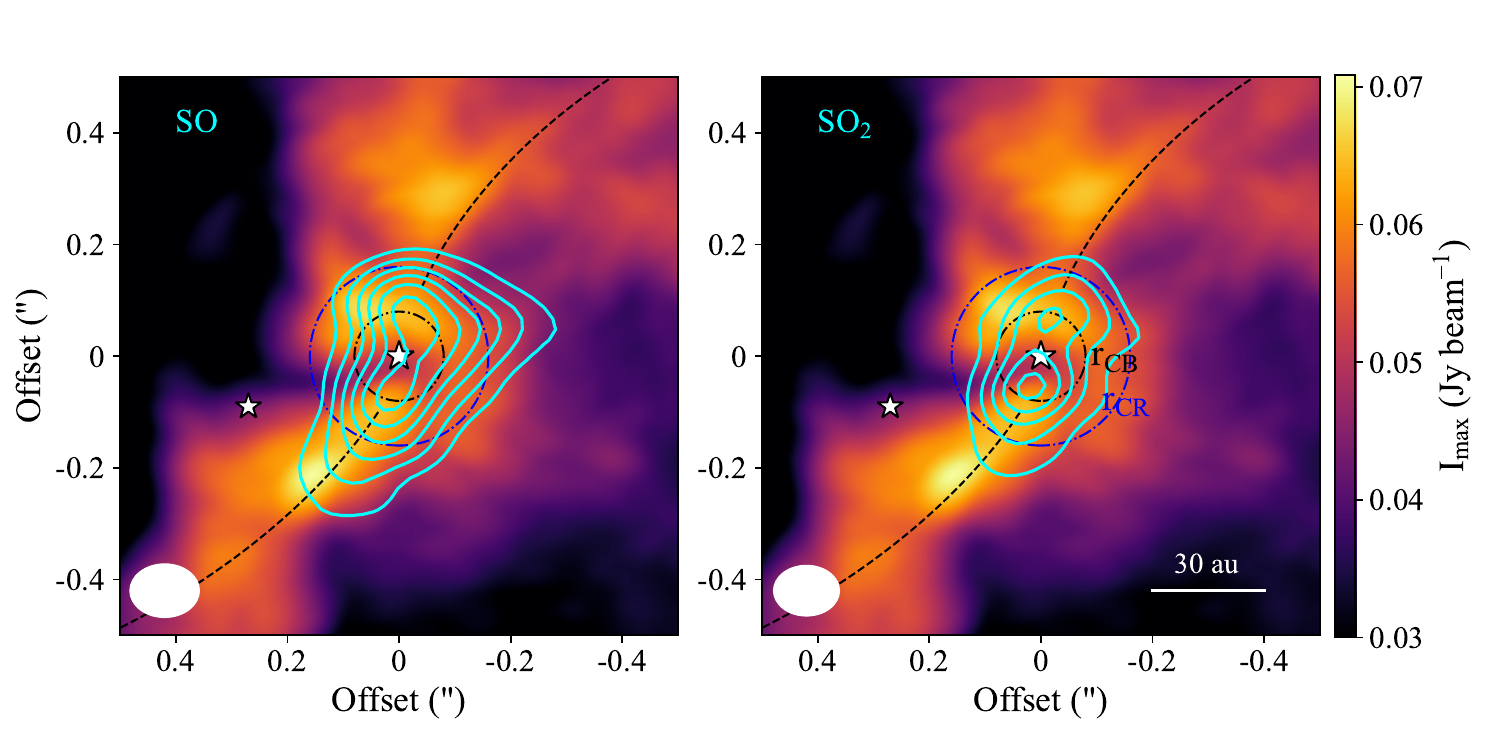}
        \caption{
        CO (moment 8) emission in colorscale superimposed with SO (\textit{left}) and SO$_{2}$ (\textit{right}) contours from moment 0 maps (in steps of 5$\sigma$). The dashed black curves indicate the direction of the proposed streamers. The dashed black and blue circles represent the extent of the centrifugal barrier and centrifugal radius, respectively, from \cite{Artur2022}. The synthesized beam is represented by a filled white ellipse. 
        }
        \label{fig:SO_moms_comb}
\end{figure*}

Position-velocity (PV) diagrams of CO 2--1 are shown in Fig.~\ref{fig:CO_PV}, for different position angles (PAs). Infalling signatures are seen in the blueshifted emission for all the PAs, and the weakest components are observed for PA~=~60$\degr$ and 80$\degr$, with negative offsets (i.e., to the east of the system). The redshifted emission is less clear than the blueshifted one, but shows an infalling profile for PA~=~160$\degr$ and the weakest emission for 60$\degr$~$\leq$~PA~$\leq$~100$\degr$. The weak emission seen in PA~=~40$\degr$ located at 2$\farcs$0 seems to trace a denser region of the outflow cavity wall. Note that the LAS of these observations is 1$\farcs$0; therefore, any extended structure beyond 1$\farcs$0 is filtered out by the interferometer.

\subsubsection{SO and SO$_{2}$ as tracers of accretion shocks} 

Figure~\ref{fig:SO_moms} shows moment 0 and 1 maps for the brightest SO transition, 7$_{8}$--6$_{7}$ (\textit{E$_\mathrm{u}$}~=~81.2~K), and SO$_{2}$ 22$_{2,20}$--22$_{1,21}$ (\textit{E$_\mathrm{u}$}~=~248.0~K). The elongated structure seen toward the southeast in SO is consistent with the shocked region proposed by \cite{Artur2022}. Additionally, it matches the direction of the SE streamer candidate, seen in CO in Fig.~\ref{fig:CO_moms}. Unlike SO, the SO$_{2}$ transition shows the peak of emission toward the south, within the continuum, but the velocity gradient also follows the direction of the SE streamer candidate. 

\cite{Artur2022} proposed that the SO$_{2}$ emission toward IRS~44 is tracing accretion shocks, given the high temperatures (above 90~K) and high densities ($\geq$10$^{8}$~cm$^{-3}$) that were estimated by analyzing six SO$_{2}$ and three $^{34}$SO$_{2}$ transitions at an angular resolution of 0$\farcs$1. They proposed that the radius of the centrifugal barrier (\textit{r$_\mathrm{CB}$}) is 0$\farcs$08, inside which the gas motion is expected to be Keplerian \citep[e.g.,][]{Sakai2014, Oya2018}. The \textit{r$_\mathrm{CB}$} is half of the centrifugal radius (\textit{r$_\mathrm{CR}$}), and infalling-rotation motions are present between both radii. Beyond \textit{r$_\mathrm{CR}$} the gas is pure infall. The extents of the proposed \textit{r$_\mathrm{CB}$} and \textit{r$_\mathrm{CR}$} are shown in Fig.~\ref{fig:SO_moms_comb}, where SO and SO$_{2}$ emission is superimposed with the CO emission. The peak of the SO and SO$_{2}$ emission is seen toward the edges of the \textit{r$_\mathrm{CB}$}, and most of the emission arises from the region where rotating-infalling motions were proposed.   

\cite{Artur2022} only targeted one SO transition; therefore, the physical properties of the SO emitting gas was estimated from the SO$_{2}$ results. Here we targeted five SO transitions, plus two $^{34}$SO lines (see Table~\ref{table:molecules}). The SO line with the lowest \textit{E$_\mathrm{u}$} (15.8~K; 1$_{0}$--0$_{1}$) is the weakest one, showing some hint of emission in its spectrum (see Fig.~\ref{fig:Spec_S} in Appendix~\ref{Ap3}), but its moment 0 map is very noisy. On the other hand, the brightest SO line toward IRS~44 is the one with the highest \textit{E$_\mathrm{u}$} value (81.2~K) and quantum numbers, 7$_{8}$--6$_{7}$. Note that the SO transition most commonly observed with ALMA in band 6 (6$_{5}$--5$_{4}$ and \textit{E$_\mathrm{u}$}~=~35.0~K; within the CO spectral setting) is clearly detected, but not the brightest one toward IRS~44. Spectra and moment 0 maps for S-bearing species are presented in Appendix~\ref{Ap3}, in Fig.~\ref{fig:Spec_S}.

\begin{table*}[h!]
        \caption{Excitation temperature and molecular column densities. }
        \label{table:rot_dig}
        \centering
        \begin{tabular}{l r c l }
                \hline\hline
                Species 			& \textit{T$_\mathrm{ex}$}	& \textit{N}							& Comment											\\
                                         		& (K)						& (cm$^{-2}$)						&													\\																																												\\
                \hline
		\multicolumn{4}{c}{\textbf{Estimated values sing RADEX}}  																				\\
               	\hline	
		SO				& 50 -- 220			& (0.06 -- 7.0)~$\times$~10$^{17}$			& This work											\\
		SO$_{2}$			& 101 -- 234			& (0.4 -- 1.8)~$\times$~10$^{17}$			& From \cite{Artur2022} $^{(a)}$							\\
                \hline
		\multicolumn{4}{c}{\textbf{Assumed temperatures (\textit{r}~=~0$\farcs$5)} }  																	\\
               	\hline
		CO				& 50 -- 200			& (2.9 -- 7.6)~$\times$~10$^{18}$			& Using $^{13}$CO 3--2 and $^{12}$C/$^{13}$C~=~69 $^{(b)}$		\\
		CS				& 50 -- 200			& (3.1 -- 4.9)~$\times$~10$^{13}$			& Using CS 7--6										\\
		H$_{2}$CO		& 50 -- 200			& (0.6 -- 2.6)~$\times$~10$^{13}$			& Using H$_{2}$CO 4$_{1,3}$--3$_{1,2}$						\\
		CH$_{3}$OH		& 100					& $\leq$~5.0~$\times$~10$^{14}$				& From \cite{Artur2019a} $^{(c)}$					\\
		OCS	 $^{(d)}$		& 50 -- 200			& $\leq$~1.4~$\times$~10$^{14}$			& Using OCS 28--27 									\\
		H$_{2}$S	$^{(d)}$	& 50 -- 200			& $\leq$~3.6~$\times$~10$^{13}$			& Using H$_{2}$S 3$_{3,0}$--3$_{2,1}$						\\
		H$_{2}$CS $^{(d)}$	& 50 -- 200			& $\leq$~3.2~$\times$~10$^{13}$			& Using H$_{2}$CS 10$_{0,10}$--9$_{0,9}$					\\
                \hline
        \end{tabular}
        \tablefoot{$^{(a)}$ Estimated from a circular region with \textit{r}~=~0$\farcs$2. $^{(b)}$ From \cite{Wilson1999}. $^{(c)}$ Estimated from a 0$\farcs$4~$\times$~0$\farcs$3 region. $^{(d)}$ Upper limits correspond to an integrated flux of 3$\sigma$.}
\end{table*}

\section{Analysis}
\subsection*{Excitation temperatures and molecular column densities} \label{Trot}
Apart from CO isotopologs, where the emission appears to be optically thick, SO is the only species with multiple transitions clearly detected. Thus, the SO physical parameters were estimated by comparing a grid of models with observed line ratios. For CO, CS, H$_{2}$CO, OCS, H$_{2}$S, and H$_{2}$CS, column densities were calculated assuming a range of temperatures, and values for SO$_{2}$ and CH$_{3}$OH were taken from the literature \citep{Artur2022,Artur2019a}. 	

\subsection{SO} 
To estimate the excitation temperature (\textit{T$_\mathrm{ex}$}) and molecular column densities of SO (\textit{N$_\mathrm{SO}$}), we employed the non-local thermodynamic equilibrium (LTE) radiative transfer code RADEX and compared the models with the observed intensity ratios between the four SO transitions clearly detected (see Table~\ref{table:molecules} and Fig.~\ref{fig:Spec_S}). Energy levels, transitions frequencies, and Einstein A coefficients were taken from the Cologne Database for Molecular Spectroscopy (CDMS), while collisional rates were taken from \cite{Lique2006}. The details of the RADEX models are presented in Appendix~\ref{Ap5}. The observed intensity ratios (see Fig.~\ref{fig:ratios}) allow us to discard \textit{n$_\mathrm{H}$}~=~10$^{6}$~cm$^{-3}$, given that there is no range of temperatures and densities that contains the observed line ratios (see Fig.~\ref{fig:radex}). For \textit{n$_\mathrm{H}$}~=~10$^{7}$~cm$^{-3}$ and \textit{n$_\mathrm{H}$}~=~10$^{8}$~cm$^{-3}$, kinetic temperatures (\textit{T$_\mathrm{kin}$}) of between 50 and 220~K are possible, while \textit{N$_\mathrm{SO}$} ranges between 0.06 and 7.0~$\times$~10$^{17}$~cm$^{-2}$ (see Fig.~\ref{fig:radex}). We note that temperatures above 220~K are not shown in Fig.~\ref{fig:radex}, but this upper limit was chosen given that gas-phase SO$_{2}$ formation is more efficient at temperatures below 200~K (see Sect.~5.1). At \textit{T$_\mathrm{kin}$}~=~300~K, \textit{n$_\mathrm{H}$} increases slightly, up to 8.0~$\times$~10$^{17}$~cm$^{-2}$. Table~\ref{table:rot_dig} shows a wide range for \textit{T$_\mathrm{kin}$} and \textit{N$_\mathrm{SO}$}, but these limits could be reduced with future observations of SO transitions with higher \textit{E$_\mathrm{u}$} values.      

\subsection{CO, CS, and H$_{2}$CO}
CO isotopologs are usually optically thick and absorption features are seen in their spectra (see Fig.~\ref{fig:Spec_CO} in the appendix); thus, \textit{N$_\mathrm{CO}$} was estimated from $^{13}$CO 3--2 and assuming a range of excitation temperatures (50 -- 200~K). Assuming LTE conditions, the following equation was employed:

\begin{equation} 
	N = \frac{N_\mathrm{u}}{g_\mathrm{u}} Q(T_\mathrm{ex}) \ \mathrm{exp}\left( \frac{E_\mathrm{u}}{T_\mathrm{ex}} \right) \ ,
    \label{eq:Eq2}
\end{equation}

\noindent where \textit{N$_\mathrm{u}$} is the column density of the upper level, \textit{g$_\mathrm{u}$} the level degeneracy, \textit{E$_\mathrm{u}$} the energy of the upper level in kelvin, \textit{N} the total column density of the molecule, and \textit{Q(T$_\mathrm{ex}$)} the partition function that depends on the excitation temperature.	

\textit{N$_\mathrm{u}$} was obtained from

\begin{equation} 
    N_\mathrm{u} = \frac{4 \pi S_{\nu} \Delta v}{A_{ij} \Omega h c} \,
    \label{eq:Eq3}
,\end{equation}

\noindent where \textit{S$_{\nu}$}.$\Delta$v is the integrated flux density, \textit{A$_{ij}$} the Einstein \textit{A} coefficient, $\Omega$ the solid angle covered by the integrated area, \textit{h} the Planck constant, and \textit{c} the speed of light. Equation~(\ref{eq:Eq3}) can be rewritten as

\begin{equation}
    N_\mathrm{u} = 2375 \times 10{^6} \ \left(\frac{S_{\nu} \Delta v}{1 \ \mathrm{Jy \ km \ s^{-1}}}\right) \ \left(\frac{1 \ \mathrm{s^{-1}}}{A_{ij}}\right) \ \left(\frac{\mathrm{arcsec{^2}}}{\mathrm{\Theta_{area}}} \right) , \
    \label{eq:Eq4}
\end{equation}

\noindent where ${\Theta}_\mathrm{area}$ is the area of integration and \textit{N$_\mathrm{u}$} was obtained in particles per square centimeter. Equations~(\ref{eq:Eq2}) and (\ref{eq:Eq4}) were used to estimate the total column density for CO, CS, and H$_{2}$CO, over a circular region with \textit{r}~=~0$\farcs$5, and assuming a range of \textit{T$_\mathrm{ex}$} between 50 and 200~K. The calculated values are listed in Table~\ref{table:rot_dig}.  	

\subsection{OCS, H$_{2}$S, and H$_{2}$CS }	
\begin{figure}[h!]
        \centering
        \includegraphics[width=.48\textwidth]{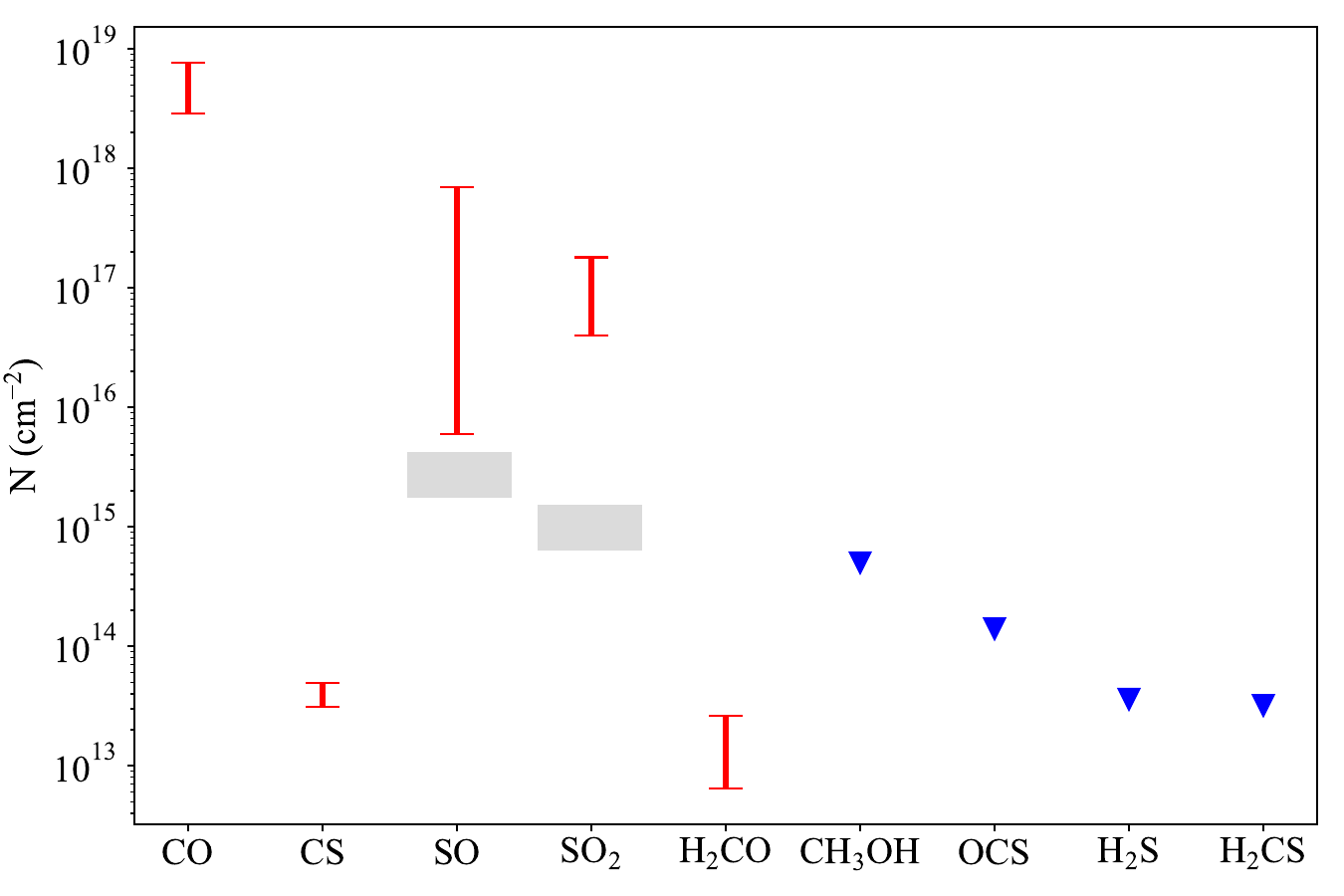}
        \caption[]{\label{fig:Nall}
        Molecular column densities. Values are taken from Table~\ref{table:rot_dig}, where red bars represent the estimated ranges and blue triangles indicate upper limits. Horizontal gray bars represent average values ($\pm$1$\sigma$) from Class I sources in the Perseus star-forming region \citep{Artur2023}. 
        }
\end{figure}

For non-detections, an integrated flux equivalent to 3$\sigma$ was employed in Equation~(\ref{eq:Eq4}), and calculated column densities should be interpreted as upper limits. In cases in which more that one transition was targeted (such as OCS and H$_{2}$S), the line with the highest \textit{A$_{ij}$} was selected to estimate the total molecular column density. Temperatures and molecular column densities are presented in Table~\ref{table:rot_dig} and Fig.~\ref{fig:Nall}.

\section{Discussion} 
In this section, we interpret the presence of bright lines from species, such as SO and SO$_{2}$, and the absence (or weak emission) of other targeted transitions.

\subsection{Bright SO and SO$_{2}$}

\cite{Artur2023} analyzed the SO and SO$_{2}$ emission of a sample of 14 Class I sources in the Perseus star-forming region, and estimated the SO and SO$_{2}$ column densities of eight and five sources, respectively. They found average values of 2.5~$\times$~10$^{15}$~cm$^{-2}$ for SO (from SO 6$_{6}$--5$_{5}$ and $^{34}$SO 5$_{6}$--4$_{5}$ multiplied by 22) and 9.9~$\times$~10$^{14}$~cm$^{-2}$ for SO$_{2}$ 14$_{0,14}$--13$_{1,13}$. The only Class I hot-corino source, Per-emb 44, was not included in this calculation, as it shows optically thick SO and SO$_{2}$ emission. Figure~\ref{fig:Nall} shows that the SO and SO$_{2}$ column densities toward IRS~44 are higher than the average values from the Perseus sources (represented by the horizontal gray bars). The lower limit estimated for \textit{N$_\mathrm{SO}$} toward IRS~44 is slightly higher than that from the Perseus sources; however, for \textit{N$_\mathrm{SO_2}$} the values toward IRS~44 are considerable higher (about two orders of magnitude) than the mean value toward the Perseus sources. These high values might be a consequence of gas-phase reactions, direct desorption from dust grains, or a combination of both processes. The main reactions for SO and SO$_{2}$ formation in the gas phase are

\begin{equation}
\begin{split}
    & \mathrm{S + OH \rightarrow SO + H}, \\
    & \mathrm{S + O_{2} \rightarrow SO + O}, \\
\end{split}
\label{eq:R1}
\end{equation}

\noindent and

\begin{equation}
    \mathrm{SO + OH \rightarrow SO_{2} + H}. \\
\label{eq:R2}
\end{equation}

Reaction~\ref{eq:R2} is more efficient at temperatures below 200~K \citep{Charnley1997}, where the presence of OH in the gas phase is required. \cite{Karska2018} did not detect OH toward IRS~44 from \textit{Herschel}/PACS observations, but infrared observations are more sensitive to optically thin and extended warm gas. Additionally, if the OH emitting region is comparable to the SO one ($\leq$0$\farcs$5), the emission will be diluted within the Herschel beam area, and therefore not detectable. Thus the presence of OH cannot be ruled out.

Another possibility for the high column densities of SO and SO$_{2}$ is the direct desorption from dust grains. When the infalling material encounters the outer regions of the disk, it generates accretion shocks, increasing the dust temperature and releasing molecules to the gas phase. The broad spectra seen for SO and SO$_{2}$ (from -15 to 15~km~s$^{-1}$) with multiple peaks at different velocities (see Fig.~\ref{fig:Spec_S}) suggest that the SO and SO$_{2}$ emission is tracing shocked regions. \cite{vanGelder2021} estimated that, for a density of 10$^{8}$~cm$^{-3}$, the sublimation temperatures of SO$_{2}$ and SO are 62~K and 37~K, respectively. Additionally, they modeled accretion shocks at the disk envelope interface and concluded that desorption of SO$_{2}$ ice is efficient in high-density environments ($\geq$10$^{8}$~cm$^{-3}$) and shock velocities of $\geq$10~km~s$^{-1}$. These values are consistent with the H$_{2}$ number density found by \cite{Artur2022}, \textit{n$_\mathrm{H}$}~$\geq$~10$^{8}$~cm$^{-3}$, and with the velocity gradient seen in Figs~\ref{fig:CO_moms} and~\ref{fig:SO_moms}. Therefore, some degree of direct SO$_{2}$ desorption is expected toward IRS~44.

\subsection{Shocks: Origins of S-bearing species and efficient production of H$_{2}$O}

Atomic sulfur is also required in the gas phase to form SO via Reaction~\ref{eq:R1}. \cite{Anderson2013} studied the S~I abundance in shocked gas with \textit{Spitzer} observations and found that atomic sulfur is a major reservoir of sulfur in shocked gas. Sulfur would be present is some form that is released from grains as atoms, perhaps via sputtering, within the shock. Once in the gas phase, atomic sulfur can be converted into a molecular form, or ionized on rapid timescales of $\sim$60/\textit{G$_\mathrm{0}$}~yr \citep[where \textit{G$_\mathrm{0}$}~=~1 is the interstellar radiation field;][]{Habing1968}. Recent works have proposed that sulfur is locked into dust grains in the form of organo-sulfur species \citep{Laas2019}, sulfide minerals such as FeS \citep{Kama2019}, sulfur chains such as S$_{8}$ \citep{Shingledecker2020, Cazaux2022}, and adsorption of S$^{+}$ followed by grain surface chemistry \citep{Fuente2023}. 

Apart from direct SO and SO$_{2}$ sublimation, \cite{vanGelder2021} propose that H$_{2}$O is efficiently produced in shocks with velocities $\geq$4~km~s$^{-1}$ through the reaction of OH with H$_{2}$. In addition, in dense environments (\textit{n$_\mathrm{H}$}~$\geq$~10$^{7}$~cm$^{-3}$), these shocks will thermally desorb H$_{2}$S for \textit{T$_\mathrm{dust}$}~=~47~K. Once in the gas phase, if the UV radiation intensity is large enough (\textit{G$_\mathrm{0}$}) $\geq$10, H$_{2}$O photodissociates into OH + H and photodissociation of H$_{2}$S significantly increases the abundance of atomic S and SH. The latter is also involved in the formation of SO and S via   

\begin{equation}
\begin{split}
    & \mathrm{SH + O \rightarrow SO + H}, \\
    & \mathrm{SH + H \rightarrow S + H_{2}}. \\
\end{split}
\label{eq:R3}
\end{equation}

\noindent Photodissociation of H$_{2}$S and H$_{2}$O will therefore increase the OH, S, and SH abundances, promoting SO and SO$_{2}$ formation in the gas phase (Reactions~\ref{eq:R1}, \ref{eq:R2}, and \ref{eq:R3}). 

The L1688 cloud in the Ophiuchus star-forming region harbors a rich cluster of young stellar objects (YSOs) at various evolutionary stages and two OB stars \citep[HD 147889 and $\rho$ Oph A;][]{Wilking1983}. The UV intensity, \textit{G$_\mathrm{0}$}, ranges between 100 and 1000 in the L1688 cloud, reaching its maximum value around the OB stars \citep[][]{Rawlings2013, Xia2022}. These values are in agreement with the condition needed to photodissociate H$_{2}$O and H$_{2}$S (i.e., \textit{G$_\mathrm{0}$} $\geq$10).

\subsection{Weak CS and H$_{2}$CO}

The main destruction path of SO is the reaction with atomic carbon \citep{Bergin1997}: 

\begin{equation}
    \mathrm{SO + C \rightarrow CS + O}, \\
\label{eq:R4}
\end{equation}

\noindent where an increase in the CS abundance is expected. However, CS is very weak around the inner regions ($\leq$50~au) of IRS~44, its emission peak is slightly offset ($\sim$0$\farcs$15 or 20~au) from the redshifted emission of SO and SO$_{2}$ (see Fig.~\ref{fig:Spec_S}), and its spectrum shows a narrow peak. This suggests that CS is not tracing the same material as SO and SO$_{2}$, but it might be present in the more quiescent and extended envelope associated with IRS~44, which is probably also filtered out by the interferometer. The weak CS in the shocked gas might imply that most of the atomic carbon is locked into CO and reaction~\ref{eq:R4} is inefficient. 

The low column density found for H$_{2}$CO reflects its weak emission, where only one transition (out of nine; see Fig.~\ref{fig:Spec_H2CO}) is clearly detected. The \textit{E$_\mathrm{u}$} values of the targeted H$_{2}$CO transitions range from 21 to 174~K, and the detected line corresponds to \textit{E$_\mathrm{u}$}~=~47.9~K. This suggests that H$_{2}$CO may be present in the cold envelope at large scales, and not in the shocked regions closer to the protostar. H$_{2}$CO can directly sublimate from dust grains for \textit{T$_\mathrm{dust}$} $\geq$~65~K \citep{vanGelder2021} but it is also formed in the gas phase for temperatures below 100~K \citep{Loomis2015,vantHoff2020,Guzman2021} via 

\begin{equation}
    \mathrm{CH_{3} + O \rightarrow H_{2}CO + H}. \\
\label{eq:R5}
\end{equation}

\noindent Independently of the formation route, the weak H$_{2}$CO emission suggests that this molecule is being destroyed around the inner regions ($\leq$50~au) of IRS~44. \cite{vanGelder2021} proposed that H$_{2}$CO is destroyed by S$^{+}$ in shocked gas, forming CO~+~H$_{2}$S$^{+}$ or SH~+~HCO$^{+}$. The less abundant isotopolog, H$^{13}$CO$^{+}$, was detected in the inner regions of IRS~44 \citep[$\leq$150~au; ][]{Artur2019a}, indicating that H$_{2}$CO destruction by S$^{+}$ might be efficient. In this scenario, the SH abundance would also increase, and Reaction~\ref{eq:R3} will once more become efficient and promote the formation of S and SO.

\subsection{Absence of OCS and CH$_{3}$OH}

\cite{Boogert2022} and \cite{Santos2024} found a good correlation between the ice abundances of OCS and CH$_{3}$OH in massive young stellar objects (MYSOs). If this correlation is also valid for low-mass YSOs, the OCS non-detection in the gas phase toward IRS 44 agrees with the absence of CH$_{3}$OH emission \citep{Artur2019a}. If OCS and CH$_{3}$OH are abundant in the ices toward IRS 44, their non-detections in the gas phase would suggest that \textit{i)} the physical conditions related to accretion shocks are not sufficient to release them to the gas phase, or \textit{ii)} they desorb from dust grains but get destroyed after sublimation. \cite{Suutarinen2014} proposed that CH$_{3}$OH is destroyed by shocks with moderate  velocities ($\geq$10~km~s$^{-1}$), where the main products are H~+~CH$_{2}$OH. In the absence of C species (other than CO), the main destruction path for OCS in the gas phase is by HCO$^{+}$ (OCS~+~HCO$^{+}$~$\rightarrow$~HOCS$^{+}$~+~CO) or by H$_{3}^{+}$ (OCS~+~H$_{3}^{+}$~$\rightarrow$~HOCS$^{+}$~+~H$_{2}$), as has been pointed out by \cite{elAkel2022}. Additional observations are needed to explain the OCS and CH$_{3}$OH non-detections.

\begin{figure*}[h!]
	\sidecaption
        \includegraphics[width=12cm]{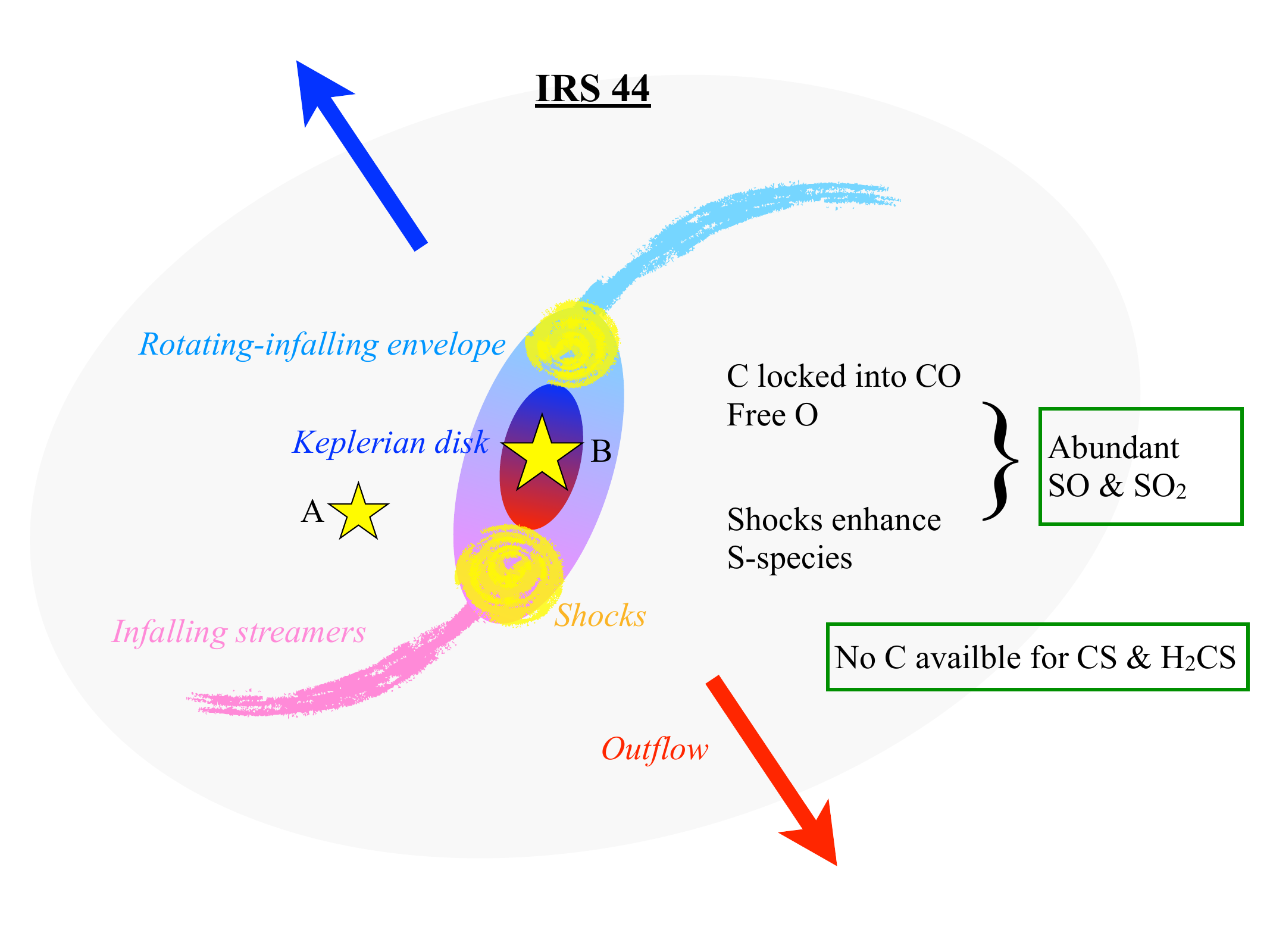}
        \caption{
        Schematic representation of the IRS~44 system. Infalling streamers produce accretion shocks in the outer regions of the rotating-infalling envelope and a Keplerian disk is expected in the innermost regions of IRS~44 B. Most of the C is locked into CO; thus, CS and H$_{2}$CS formation is not efficient. Accretion shocks sublimate S-bearing species and enhance their gas abundances,  promoting a lukewarm chemistry ($\sim$200~K) in the gas phase, mainly involving O- and S-bearing species.   
        }
        \label{fig:cartoon}
\end{figure*}

\subsection{Very low CS/SO}

The CS/SO column density ratio has been shown to be sensitive to the overall C/O ratio in the gas in Class II disks \citep[i.e., ][]{Semenov2018, LeGal2021}. In IRS 44, the low CS/SO ratio ($\leq$~0.008) is a consequence of the weak CS emission. As is shown in Fig.~\ref{fig:Spec_S} in the appendix, the CS emission peaks toward the south of IRS 44 B, slightly offset ($\sim$0$\farcs$15 or 20~au) from the redshifted peak of SO and SO$_{2}$, and $^{13}$CS and C$^{34}$S are not detected, ruling out the possibility that the CS 7--6 transition is optically thick. The low CS/SO ratio is consistent with the values toward IRS 48 \citep[CS/SO < 0.01; ][]{Booth2024}, but in high contrast with C/O ratios estimated for Class II disks \citep[$\geq$2.0;][]{Bosman2021, LeGal2021}. CS/SO ratios toward younger and more embedded Class I disks are unknown in general, given the challenge of distinguishing between the different components of the system. For example, CS is commonly tracing the disk and also more extended emission from the outflow and envelope components \citep{Artur2023}. Thus, high-angular-resolution observations and multiple transitions are essential for an accurate CS/SO estimation toward embedded sources.

Another Class I source with bright SO and SO$_{2}$ emission, but no detection of CS, is Per-emb 50 \citep{Artur2023}. This is the only source within a sample of 50 Class 0/I YSOs that shows this behavior and it has been associated with accretion shocks \citep{Zhang2023} and an infalling streamer \citep{Valdivia2022}. This may suggests that the CS intensity anticorrelates with the presence of accretion shocks.

The fact that CS and sometimes H$_{2}$CS are the main sulfur-bearing molecules detected in Class II disks \citep{LeGal2019, Law2025} may suggest that H$_{2}$CS in Class II disks is likely forming in the gas phase at low temperatures \citep[<100~K; ][]{Booth2024}, while SO and SO$_{2}$ are detected in those sources where infalling material is still significant \citep[e.g., ][]{Garufi2022} or the UV radiation can heat the dust to temperatures where SO and SO$_{2}$ can thermally desorb \citep{Booth2021}. From the discussion above, the physical conditions that seem to prevail toward the inner regions ($\leq$0$\farcs$3 or 40~au) of IRS~44 are the following: \textit{T$_\mathrm{gas}$} between 100 and 300~K, \textit{n$_\mathrm{H}$} $\geq$ 10$^{7}$~cm$^{-3}$, a shock velocity of $\geq$ 4~km~s$^{-1}$, and \textit{G$_\mathrm{0}$} $\geq$ 10. 

Figure~\ref{fig:cartoon} shows a schematic representation of the inner regions of IRS~44 and our interpretation of the data. Infalling streamers are producing accretion shocks when they encounter the outer regions of the infalling-rotating envelope. These shocks heat the dust and efficiently release S-bearing species (such as S, H$_{2}$S, SO, and SO$_{2}$), as well as promoting a lukewarm chemistry ($\sim$200~K) in the gas phase. Most of the atomic carbon might be locked into CO, so there is no available C to form CS and H$_{2}$CS, which leaves an oxygen-rich environment. The high column densities of SO and SO$_{2}$ might therefore be a consequence of both processes: direct thermal desorption from dust grains and gas-phase formation due to the availability of O- and S-bearing species. Finally, a Keplerian disk is expected in the innermost regions associated with IRS~44 B ($\leq$0$\farcs$08 or 11~au).

\section{Summary} 
This work presents new ALMA data toward the Class I source IRS 44, where multiple molecular transitions are targeted, sulfur-bearing species being the most relevant ones. The main results are summarized below:
\begin{itemize}
	\item The continuum emission shows a binary component, which is brighter in ALMA band 6 (233.0~GHz). The binarity of IRS 44 was proposed before from infrared observations; nevertheless, this is the first time that the binary components have been resolved in the sub-millimeter regime. The average total disk masses (dust + gas) calculated for source B and source A are 5.1~$\times$~10$^{-3}$ and 0.53~$\times$~10$^{-3}$~M$_{\odot}$, respectively, with one order of magnitude difference between them. 
	\item Among the CO-bearing species, $^{12}$CO, $^{13}$CO, and C$^{18}$O transitions are detected; however, only one H$_{2}$CO line is clearly detected at small scales. Colder H$_{2}$CO might be more abundant at larger and more extended scales. 
	\item For S-bearing species, six SO transitions are clearly detected (including two $^{34}$SO), while only the most abundant isotopologs of CS and SO$_{2}$ are detected. The targeted OCS, H$_{2}$S, and H$_{2}$CS transitions are not detected in this dataset. 
	\item The envelope around IRS 44 shows infalling signatures (mainly seen in CO emission) and, as the gas approaches the disk, an S shape stands out in the velocity profile of CO. This S shape, or arc-like structures, represent our streamer candidates and the direction of the southeast one agrees with the slightly extended redshifted emission seen in SO. The infalling streamers would generate accretion shocks when the material encounters the outer regions of the rotating-infalling envelope.
	\item We propose that the high SO and SO$_{2}$ column densities are driven by two main physical mechanisms: direct thermal desorption from dust grains due to the accretion shocks and gas-phase formation within the O-rich gaseous environment present in the inner regions of IRS~44~B.  
	\item Accretion shocks seem to be essential to increase the gas abundance of S, H$_{2}$S, SO, and SO$_{2}$, as well as forming H$_{2}$O in the gas phase. Later on, atomic sulfur can be ionized, while H$_{2}$S, and H$_{2}$O could be photodissociated, increasing the abundance of S$^{+}$, S, SH, O, and OH; the main reactants of SO and SO$_{2}$.      
	\item The weak CS emission will be explained by the lack of atomic C available in the gas, where CO formation depletes the available carbon and an oxygen-rich chemistry is promoted. H$_{2}$CO might be destroyed in the shocked gas by S$^{+}$, which explains its weak emission. Nevertheless, both CS and H$_{2}$CO are expected to be abundant in the colder and more extended envelope around the IRS~44 system.    
	\item The absence of OCS is in agreement with the non-detection of CH$_{3}$OH, as both species seem to share a common origin in ice mantles. If OCS and CH$_{3}$OH are abundant in the ices toward IRS 44, those species remain locked and the physical conditions related to accretion shocks are not sufficient to release them to the gas phase, or they are both thermally desorbed and later on destroyed in the gas phase.    
	\item Weak CS emission, absence of COMs, and bright SO and SO$_{2}$ emission seem to be a good recipe for accretion shocks with moderate velocities ($\sim$10~km~s$^{-1}$) toward embedded sources. 
	\item We propose that the physical conditions toward the inner regions ($\leq$0$\farcs$3 or 40~au) of IRS~44 are the following: \textit{T$_\mathrm{gas}$} between 100 and 300~K, \textit{n$_\mathrm{H}$} $\geq$ 10$^{7}$~cm$^{-3}$, a shock velocity of $\geq$~4~km~s$^{-1}$, and \textit{G$_\mathrm{0}$}~$\geq$~10.   

\end{itemize}

IRS 44 is an ideal candidate with which to study the chemical consequences of accretion shocks and the complex dynamics of young and embedded proto-binary systems. Future observations of H$_{2}$$^{18}$O, HDO, and OH would allow us to test our proposed scenario, while continuum ALMA observations of the dust with a higher angular resolution ($\leq$0$\farcs$1) and in multiple bands would allow us to better constrain the disk's physical properties and the dust temperature, which is expected to increase in shocked regions. There are other S-bearing species that are not targeted in these observations, such as CCS, NS, NS$^{+}$, and salts such as NH$_{4}$SH \citep[e.g., ][]{Slavicinska2025}. Given the weak CS emission and non-detection of H$_{2}$CS, CCS is not expected to be abundant, however, N-bearing species containing sulfur and salts could constitute an important sulfur sink. Future estimations of the CS/SO ratio toward a larger sample of more embedded disks would provide valuable information on the material transport and the chemical evolution from younger to more evolved disks.

\begin{acknowledgements}

We thank the anonymous referee for a number of good suggestions that helped us to improve this work. This paper makes use of the following ALMA data: ADS/JAO.ALMA$\#$2022.0.00209.S and ADS/JAO.ALMA$\#$2019.1.01792.S. ALMA is a partnership of ESO (representing its member states), NSF (USA) and NINS (Japan), together with NRC (Canada), MOST and ASIAA (Taiwan), and KASI (Republic of Korea), in cooperation with the Republic of Chile. The Joint ALMA Observatory is operated by ESO, AUI/NRAO and NAOJ. The National Radio Astronomy Observatory is a facility of the National Science Foundation operated under cooperative agreement by Associated Universities, Inc. V.V.G. acknowledge support from the ANID -- Millennium Science Initiative Program -- Center Code NCN2024\_001, from FONDECYT Regular 1221352, and ANID CATA-BASAL project FB210003. D.H. is supported by the Ministry of Education of Taiwan (Center for Informatics and Computation in Astronomy grant and grant number 110J0353I9) and the National Science and Technology Council, Taiwan (Grant NSTC111-2112-M-007-014-MY3, NSTC113-2639-M-A49-002-ASP, and NSTC113-2112-M-007-027).

\end{acknowledgements}

\bibliographystyle{aa} 
\bibliography{References} 

\begin{appendix}

\section{Molecular transitions} \label{Ap0}
The targeted molecular transitions within three different spectral setting are present in Table~\ref{table:molecules}, where the detected lines are listed at the beginning.

\begin{table*}[h!]
        \caption{Spectral setup, parameters of the observed molecular transitions, and integrated line fluxes. }
        \label{table:molecules}
        \centering
        \begin{tabular}{l l l l c c c}
                \hline\hline
                Species 			& Transition                            			& Frequency  	& \textit{E$_\mathrm{u}$}      	& \textit{A$_{ij}$}	& g$_{u}$		& \textit{S$_\mathrm{\nu}$}.$\Delta$v $^{(a)}$ 	\\
                                         		&               						& (GHz)           	& (K)                                   	& (s$^{-1}$)		& 			& (Jy~km~s$^{-1}$)	\\
                \hline		
                \multicolumn{7}{c}{\textbf{Detected}}  																												\\
                \hline
		CO				& 2--1							& 230.5380	& 16.6					& 6.9e-7			& 5			& 20.415~$\pm$~0.014	 		\\
		$^{13}$CO		& 2--1							& 220.3987	& 15.9					& 3.0e-7			& 10			& 1.532~$\pm$~0.012			\\	
		$^{13}$CO		& 3--2							& 330.5880	& 31.7					& 2.2e-6			& 7			& 7.201~$\pm$~0.036			\\
		C$^{18}$O		& 2--1							& 219.5604	& 15.8					& 6.0e-7			& 5			& 0.071~$\pm$~0.012			\\
		C$^{18}$O		& 3--2							& 329.3305	& 31.6					& 2.2e-6			& 7			& 0.682~$\pm$~0.022	 		\\
		CS				& 7--6							& 342.8829	& 65.8					& 8.4e-4			& 15			& 1.027~$\pm$~0.020			\\
		SO				& 6$_{5}$--5$_{4}$					& 219.9494	& 35.0					& 1.3e-4			& 13			& 3.258~$\pm$~0.010			\\
		SO				& 7$_{7}$--6$_{6}$					& 301.2861	& 71.0					& 3.4e-4			& 15			& 6.861~$\pm$~0.026 			\\
		SO				& 1$_{2}$--0$_{1}$					& 329.3855	& 15.8					& 1.4e-5			& 3			& 0.399~$\pm$~0.042			\\
		SO				& 3$_{3}$--3$_{2}$					& 339.3415	& 25.5					& 1.5e-5			& 7			& 0.263~$\pm$~0.011			\\
		SO				& 7$_{8}$--6$_{7}$					& 340.7142	& 81.2					& 5.0e-4			& 15			& 9.315~$\pm$~0.045 			\\
		$^{34}$SO		& 6$_{7}$--5$_{6}$					& 290.5622	& 63.8					& 3.0e-4			& 13			& 0.586~$\pm$~0.027 			\\
		$^{34}$SO		& 9$_{8}$--8$_{7}$					& 339.8573	& 77.3					& 5.1e-4			& 19			& 0.755~$\pm$~0.014 			\\
		SO$_{2}$			& 22$_{2,20}$--22$_{1,21}$			& 216.6433	& 248.0					& 9.3e-5			& 45			& 0.762~$\pm$~0.008			\\		
		H$_{2}$CO		& 4$_{1,3}$--3$_{1,2}$				& 300.8366	& 47.9					& 7.2e-4			& 27			& 0.099~$\pm$~0.008			\\
		\hline
		\multicolumn{7}{c}{\textbf{Non-detected}}  																												\\
               	\hline	
		$^{13}$CS		& 5--4							& 231.2207	& 33.3					& 2.5e-4			& 22			& $\leq$~0.009		 	\\
		C$^{34}$S		& 6--5							& 289.2091	& 38.2					& 4.8e-4			& 13			& $\leq$~0.027			\\
		$^{34}$SO$_{2}$	& 21$_{7,15}$--22$_{6,16}$			& 216.5935	& 328.0					& 2.0e-5			& 43			& $\leq$~0.030			\\
		$^{34}$SO$_{2}$	& 5$_{5,1}$--6$_{4,2}$				& 329.4994	& 73.1					& 8.5e-6			& 11			& $\leq$~0.075	 		\\
		$^{34}$SO$_{2}$	& 5$_{3,3}$--4$_{2,2}$				& 342.2089	& 35.1					& 3.1e-4			& 11			& $\leq$~0.036			\\
		$^{34}$SO$_{2}$	& 20$_{1,19}$--19$_{2,18}$			& 342.2316	& 199.0					& 3.1e-4			& 41			& $\leq$~0.036			\\
		$^{34}$SO$_{2}$	& 12$_{4,8}$--12$_{3,9}$				& 342.3320	& 110.0					& 3.1e-4			& 25			& $\leq$~0.036	 		\\
		OCS				& 19--18							& 231.0610	& 111.0					& 3.6e-5			& 39			& $\leq$~0.006			\\
		OCS				& 24--23							& 291.8397	& 175.0					& 7.2e-5			& 49			& $\leq$~0.022			\\
		OCS				& 28--27							& 340.4492	& 237.0					& 1.2e-4			& 57			& $\leq$~0.021			\\
		O$^{13}$CS		& 19--18							& 230.3175	& 111.0					& 3.5e-5			& 39			& $\leq$~0.009			\\
		H$_{2}$S			& 2$_{2,0}$--2$_{1,1}$				& 216.7104	& 84.0					& 4.9e-5			& 5			& $\leq$~0.009			\\
		H$_{2}$S			& 3$_{3,0}$--3$_{2,1}$				& 300.5056	& 169.0					& 1.0e-4			& 21			& $\leq$~0.029			\\
		H$_{2}$CS		& 10$_{0,10}$--9$_{0,9}$				& 342.9464	& 90.6					& 6.1e-4			& 21			& $\leq$~0.033			\\
		H$_{2}$CO		& 9$_{1,8}$--9$_{1,9}$				& 216.5687	& 174.0					& 7.2e-6			& 57			& $\leq$~0.033			\\
		H$_{2}$CO		& 3$_{0,3}$--2$_{0,2}$				& 218.2222	& 21.0					& 2.8e-4			& 7			& $\leq$~0.006			\\
		H$_{2}$CO		& 3$_{2,1}$--2$_{2,0}$				& 218.7601	& 68.1					& 1.6e-4			& 7			& $\leq$~0.006			\\
		H$_{2}$CO		& 4$_{0,4}$--3$_{0,3}$				& 290.6234	& 34.9					& 6.9e-4			& 9			& $\leq$~0.033	 		\\
		H$_{2}$CO		& 4$_{2,3}$--3$_{2,2}$				& 291.2378	& 82.1					& 5.2e-4			& 9			& $\leq$~0.024			\\
		H$_{2}$CO		& 4$_{3,2}$--3$_{3,1}$				& 291.3804	& 141.0					& 3.0e-4			& 27			& $\leq$~0.024			\\
		H$_{2}$CO		& 4$_{3,1}$--3$_{3,0}$				& 291.3844	& 141.0					& 3.0e-4			& 27			& $\leq$~0.015			\\
		H$_{2}$CO		& 4$_{2,2}$--3$_{2,1}$				& 291.9481	& 82.1					& 5.2e-4			& 9			& $\leq$~0.027			\\
		H$_{2}^{13}$CO	& 11$_{2,9}$--12$_{0,12}$			& 231.2460	& 274.0					& 3.6e-7			& 23			& $\leq$~0.012			\\
		H$_{2}^{13}$CO	& 11$_{1,10}$--11$_{1,11}$			& 301.1541	& 243.0					& 1.3e-5			& 69			& $\leq$~0.033			\\
		H$_{2}$CN		& 9$_{1,8}$--9$_{1,9}$ 21/2--21/2 1--1	& 220.3727	& 175.0					& 8.9e-6			& 22			& $\leq$~0.006		 	\\
		c-C$_{3}$H$_{2}$	& 11$_{7,5}$--10$_{10,0}$			& 216.5519	& 192.0					& 5.3e-8			& 23			& $\leq$~0.009 			\\
		\hline
		\multicolumn{7}{c}{\textbf{Blended with SO and $^{13}$CO}}  																								\\
               	\hline
		H$_{2}$CN		& 3$_{0,3}$--2$_{0,2}$ 9/2--7/2 4--4		& 219.9391	& 21.1					& 2.9e-4			& 10			& 			 				\\
		H$_{2}$CN		& 3$_{2,2}$--2$_{2,1}$ 7/2--5/2 9--9		& 220.3875	& 68.7					& 1.5e-4			& 8			& 							\\
		H$_{2}$CN		& 3$_{2,2}$--2$_{2,1}$ 9/2--7/2 4--4		& 220.4049	& 68.6					& 1.8e-4			& 10			& 							\\
                \hline
        \end{tabular}
        \tablefoot{Detections are defined when the integrated emission in a circular area with radius of 0$\farcs$5 is above a 3$\sigma$ value.$^{(a)}$Upper limits represent a 3$\sigma$ value. $\Delta$v is 50~km~s$^{-1}$ and 20~km~s$^{-1}$ for detections and non-detections, respectively. 
			}
\end{table*}

\section{CO 2--1 from archival data} \label{Ap1}
\begin{figure}[h!]
        \centering
        \includegraphics[width=.48\textwidth]{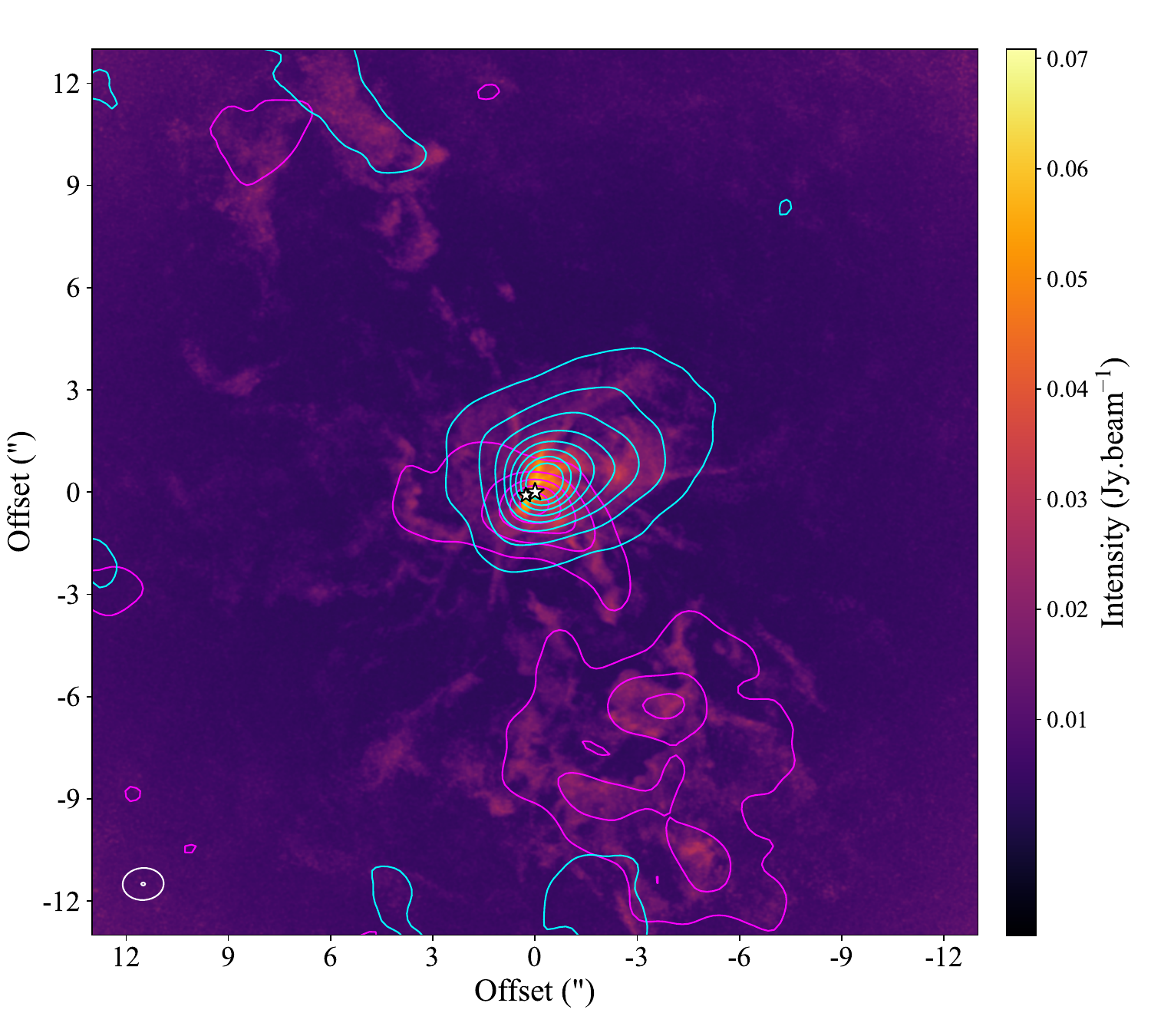}
        \caption[]{\label{fig:CO_mom8_archive}
        CO 2--1 emission at large scales. The colorscale represents the moment 8 map from this dataset, and the magenta and cyan contours show red- and blueshifted emission from archival data with poorer angular resolution and larger LAS. Magenta and cyan contours start at 60$\sigma$ (1$\sigma$~=~16~mJy~beam$^{-1}$~km~s$^{-1}$) and follow an increase of 125$\sigma$. The yellow stars represent the position of the A and B components (see Fig.~\ref{fig:cont}) and the synthesized beams are represented by white ellipses in the lower-left corner of the map.   
        }
\end{figure}

Given that our observations filtered-out emission more extended than 1$\farcs$0, we use archival data (project 2019.1.01792.S) that targeted the same CO 2--1 transition, to trace intermediate scales: from 1$\farcs$0 to 5$\farcs$0. This CO emission is presented in Fig.~\ref{fig:CO_mom8_archive} as contour maps, showing that the infalling envelope covers an angular extent of $\sim$6$\farcs$0 ($\sim$800~au), and the blueshifted emission is brighter and more extended than the redshifted one. Additionally, weaker outflow components are also seen, where blue- and redshifted emission overlap in the northeast and southwest directions, supporting the interpretation that the disk is roughly edge-on and the outflow axis lies close to the plane of the sky.

\section{Spectra}
\subsection{CO isotopologs} \label{Ap2}
\begin{figure}[h!]
        \centering
        \includegraphics[width=.24\textwidth]{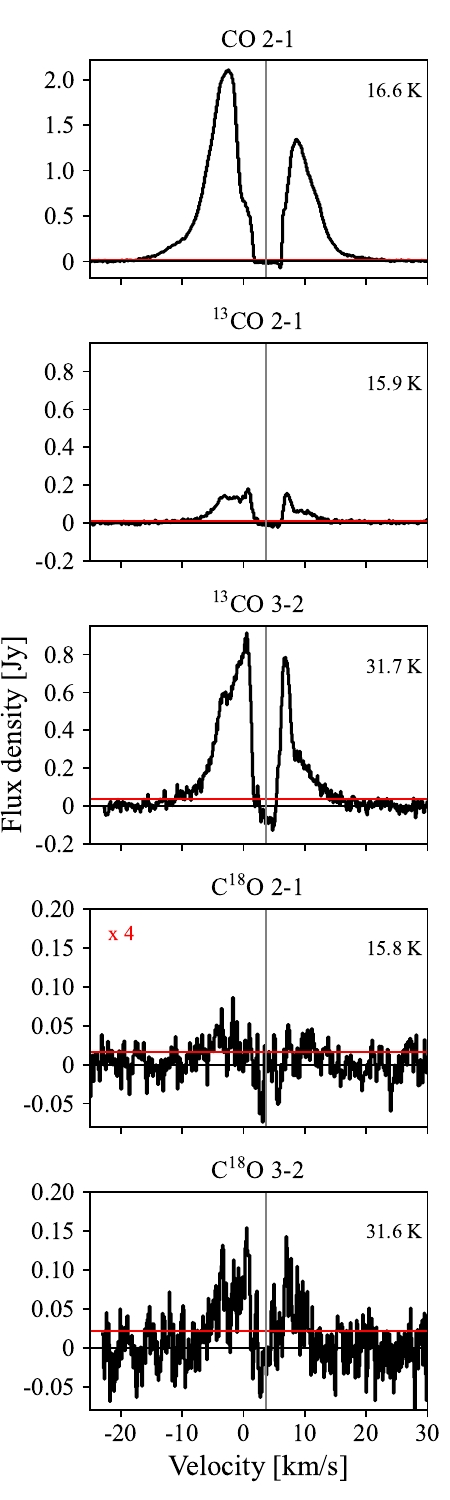}
        \includegraphics[width=.24\textwidth]{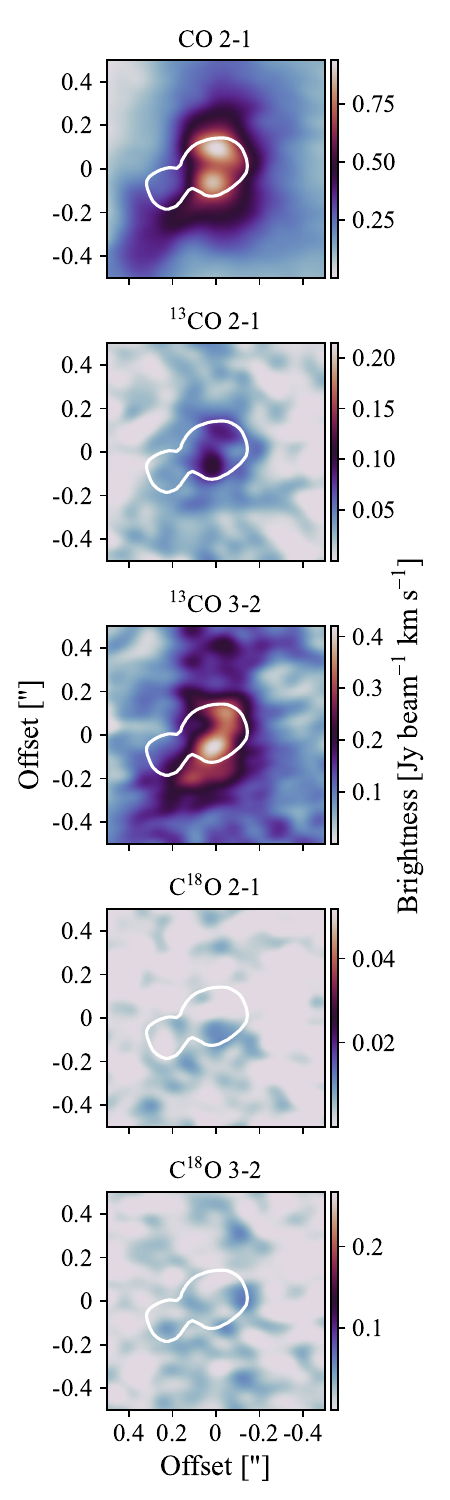}
        \caption[]{\label{fig:Spec_CO}
        Emission of CO isotopologs. \textit{Left}: Spectra taken from a region with a radius of 0$\farcs$5 around the brightest component, the upper level energy of each transition is shown in the upper-right corner of each spectrum, and the horizontal red line represents 1$\sigma$. The C$^{18}$O 2--1 spectrum has been multiplied by a factor of 4 for a better comparison. \textit{Right}: Moment 0 maps integrated from -25 to +25~km~s$^{-1}$, with respect to the systemic velocity (3.7~km~s$^{-1}$). White contours represent the continuum emission at 233~GHz at a 10$\sigma$ value. 
        }
\end{figure}

All the targeted CO isotopolog transitions are detected, where the spectra and moment 0 maps are shown in Fig.~\ref{fig:Spec_CO}. For $^{13}$CO and C$^{18}$O, two transitions are present in our spectral settings and, in both cases, the 3--2 transition is brighter than the 2--1 one. This suggests that the CO emitting region closer to the binary system is associated with more lukewarm temperatures ($\geq$30~K). Furthermore, absorption features are seen toward the five spectra at the source velocity, indicating that extended emission is being filtered-out by the interferometer. Future observations of warmer CO transitions, such as 4--3 in band 8, would provide a better estimation of the gas temperature at disk scales.

\subsection{S-bearing species} \label{Ap3}
\begin{figure*}[h!]
        \centering
        \includegraphics[width=.24\textwidth]{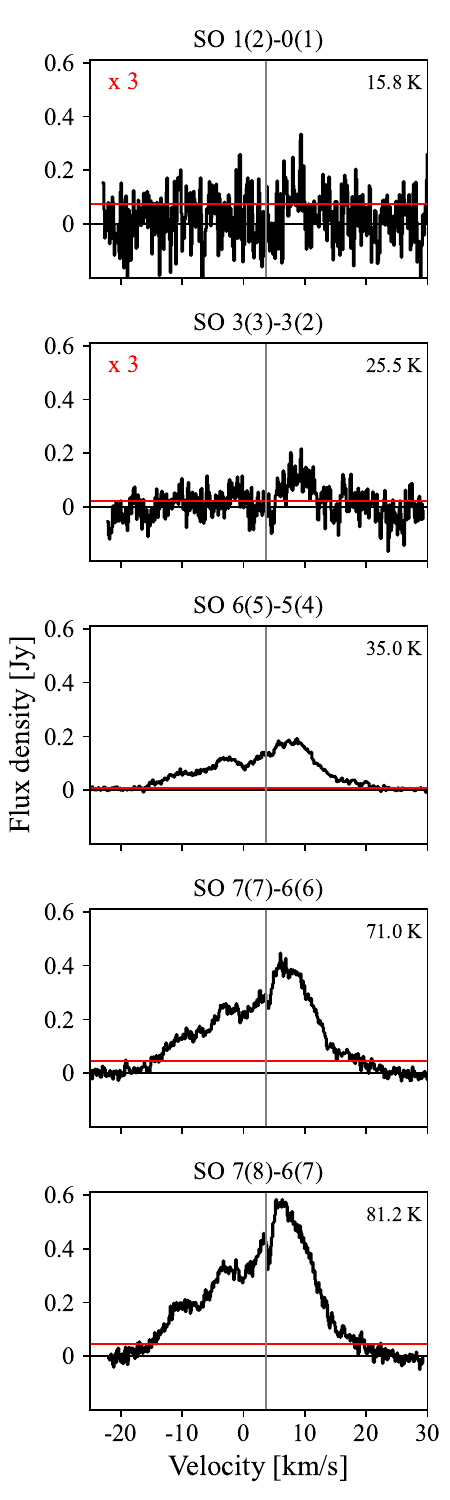}
        \includegraphics[width=.24 \textwidth]{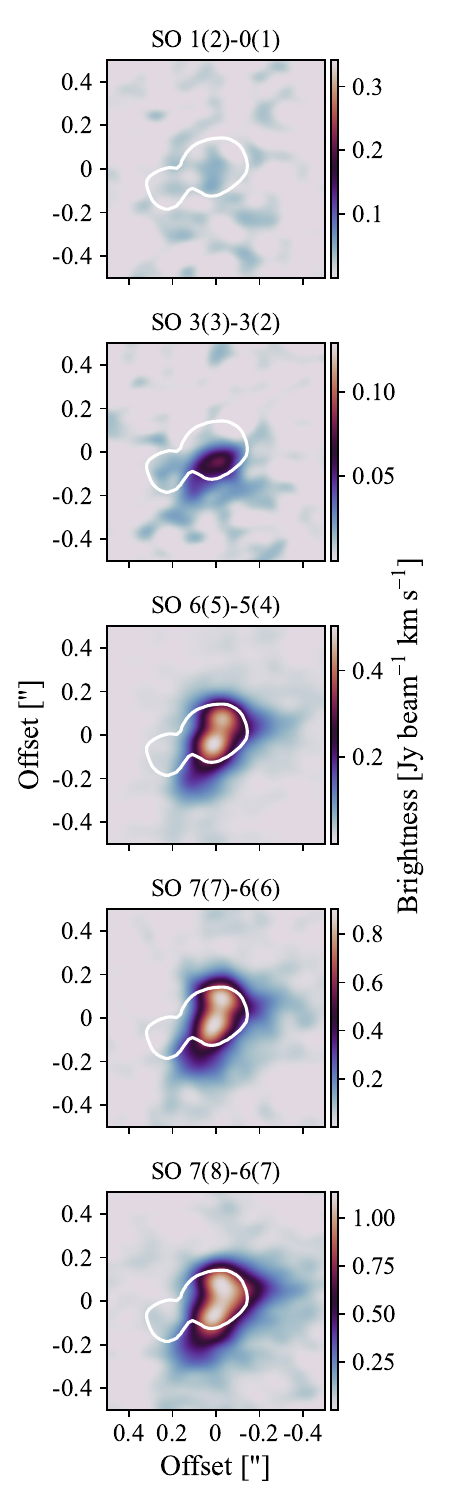}
        \includegraphics[width=.24\textwidth]{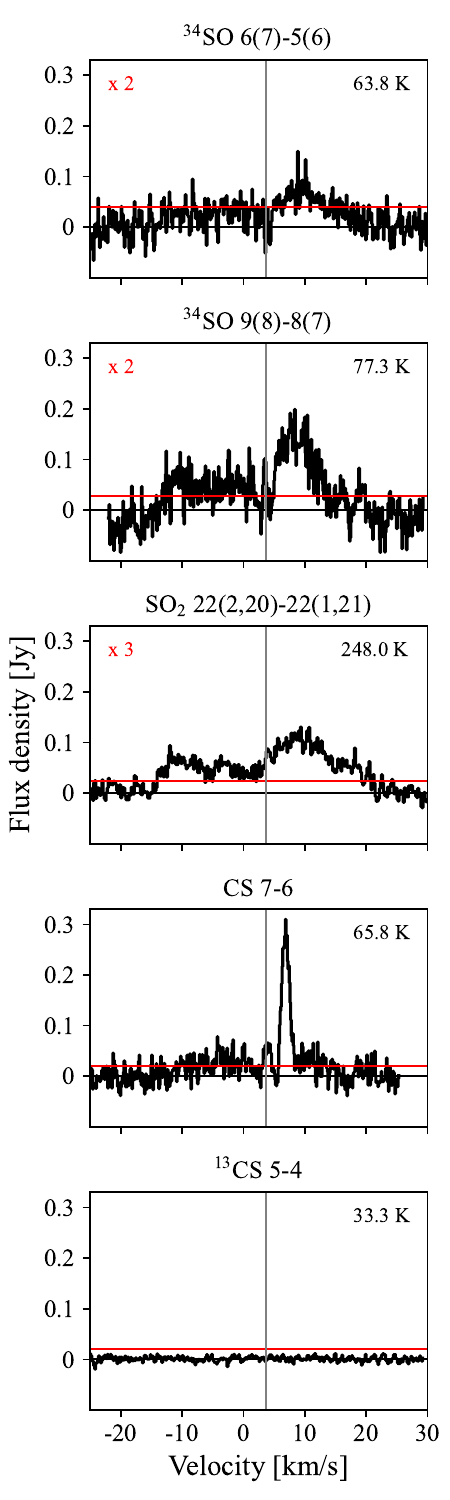}
        \includegraphics[width=.24 \textwidth]{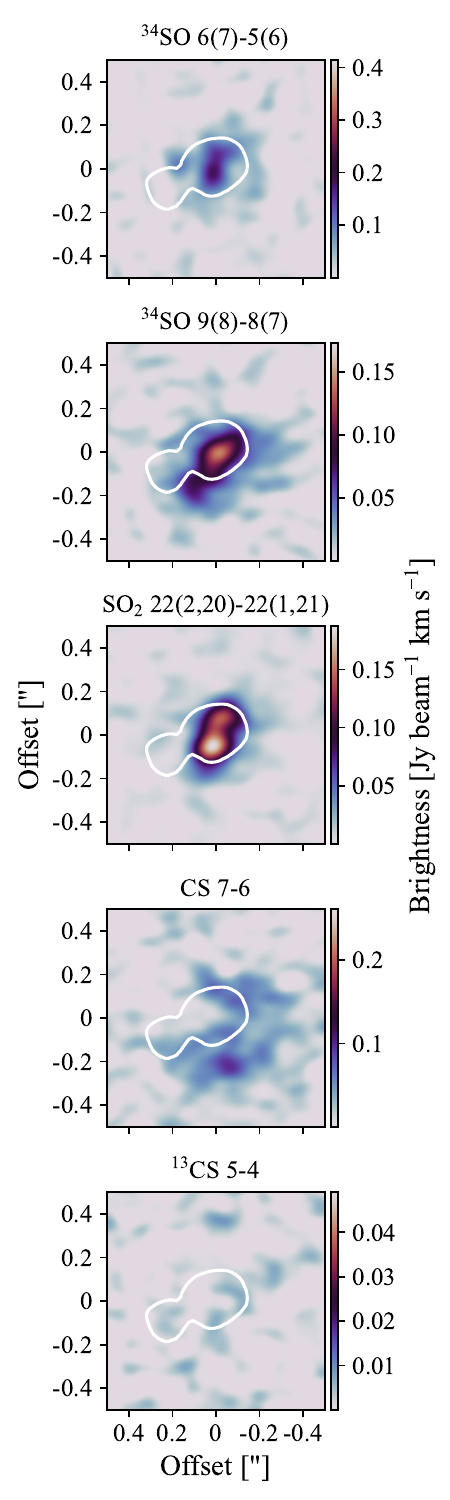}
        \caption[]{\label{fig:Spec_S}
        Same as Fig.~\ref{fig:Spec_CO} for SO, SO$_{2}$, and CS isotopologs. Weak lines, such as SO~1$_{2}$--0$_{1}$, SO~3$_{3}$--3$_{2}$, $^{34}$SO~6$_{7}$--5$_{6}$, $^{34}$SO~9$_{8}$--8$_{7}$, and SO$_{2}$~22$_{2,20}$--22$_{1,21}$, have been multiplied by a factor of 2 or 3 for a better comparison. $^{13}$CS~5--4 is not detected in these data. 
        }
\end{figure*}

Similarly to CO, spectra and moment 0 maps of SO, SO$_{2}$ and CS isotopologs are shown in Fig.~\ref{fig:Spec_S}. The brightest SO transition is the one with the highest \textit{E$_\mathrm{u}$} value (81.2~K), and all the spectra (with the exception of CS and $^{13}$CS) show similar profiles: a broad spectrum with brighter redshifted emission. The CS spectrum presents a bright and narrow redshifted peak. Although weak CS emission is seen at small scales, the integrated emission peaks toward the south of IRS 44 B, where SO and SO$_{2}$ emission is significantly weak.

\subsection{H$_{2}$CO} \label{Ap4}
Among the nine H$_{2}$CO transitions, only one of them (4$_{1,3}$--3$_{1,2}$) is clearly detected, with an absorption and emission features at redshifted velocities. H$_{2}$CO 3$_{0,3}$--2$_{0,2}$ and 4$_{0,4}$--3$_{0,3}$, the coldest transitions, present some hint of absorption, but this remains unclear.

\begin{figure*}[h!]
	\sidecaption
        \includegraphics[width=12cm]{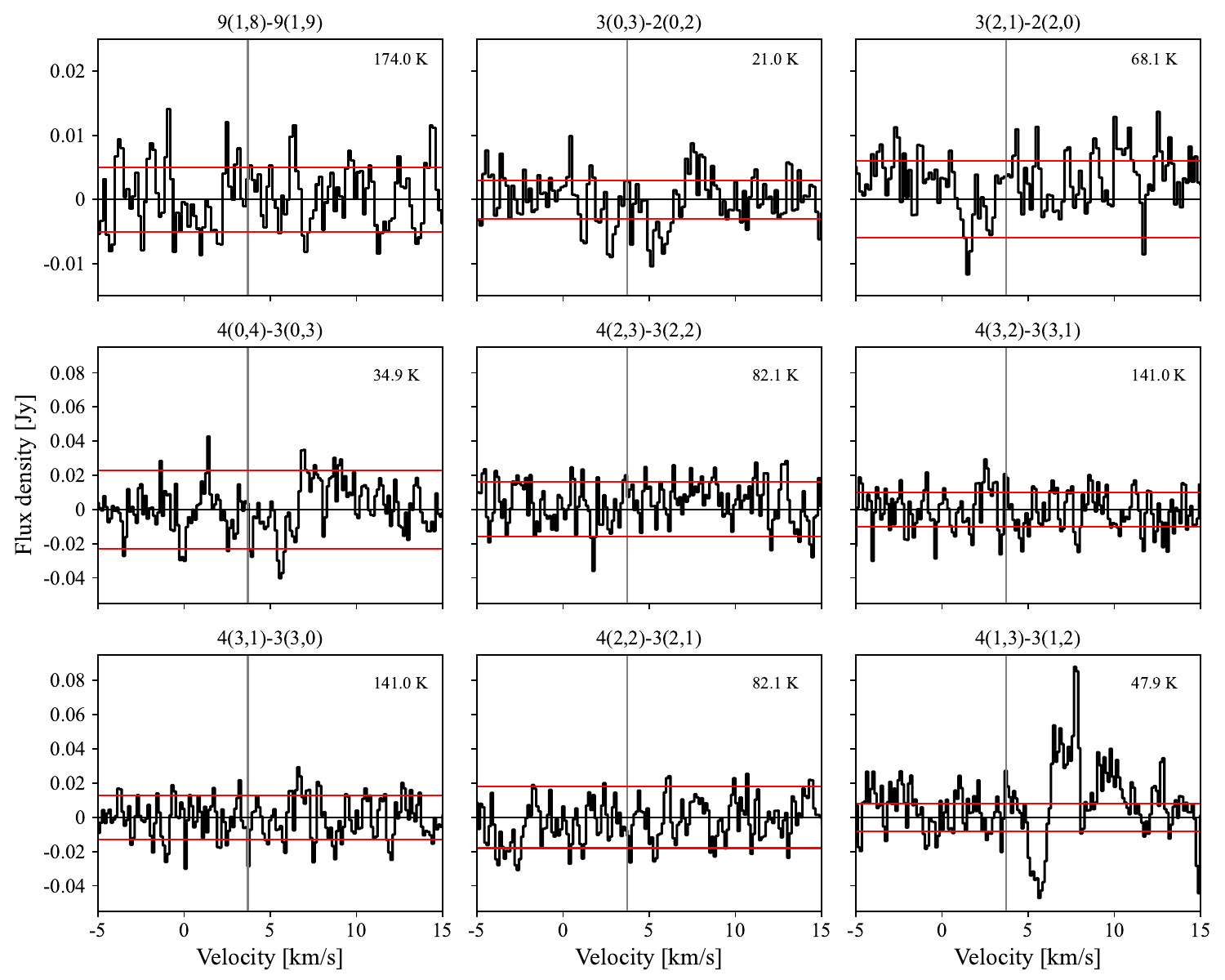}
        \caption{        
        H$_{2}$CO spectra taken from a region with a radius of 0$\farcs$5 around the brightest component. The horizontal red lines represent $\pm$1$\sigma$ values, while the vertical gray line shows the source velocity of 3.7~km~s$^{-1}$. The upper level energy of each transition is shown in the upper-right corner of each spectrum. The only clear H$_{2}$CO detection is the one that corresponds to the 4$_{1,3}$--3$_{1,2}$ transition. 
                        }
	\label{fig:Spec_H2CO}
\end{figure*}

\section{Non-LTE radiative transfer models} \label{Ap5}
\begin{figure*}[h!]
	\sidecaption
        \includegraphics[width=12cm]{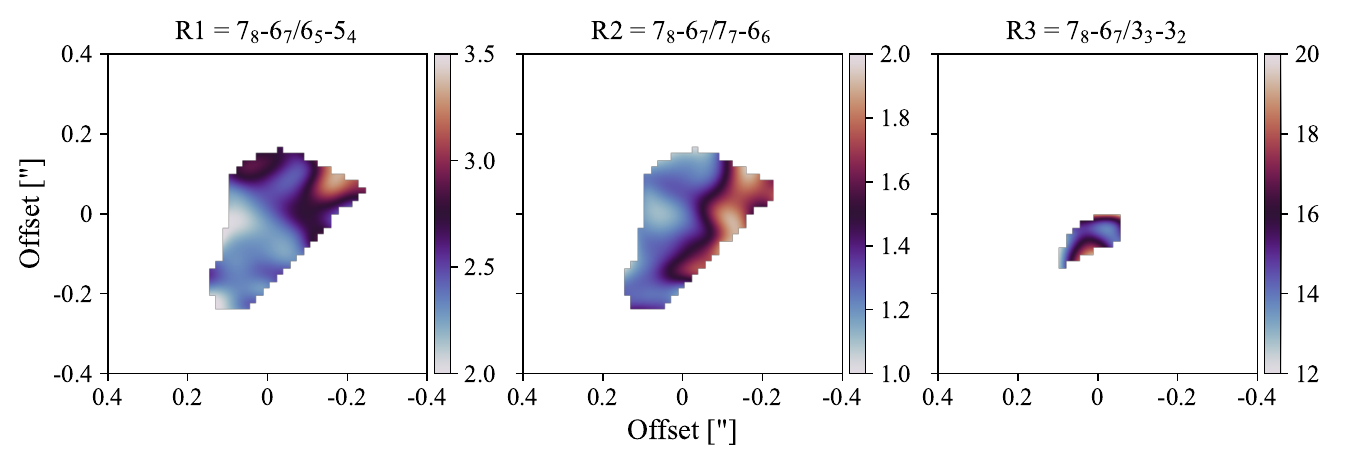}
        \caption{
	Observed intensity ratios between SO 7$_{8}$--6$_{7}$ and the other three bright SO transitions, above a 3$\sigma$ level.  
        }
        \label{fig:ratios}
\end{figure*}

\begin{figure*}[h!]
	\sidecaption
        \includegraphics[width=12cm]{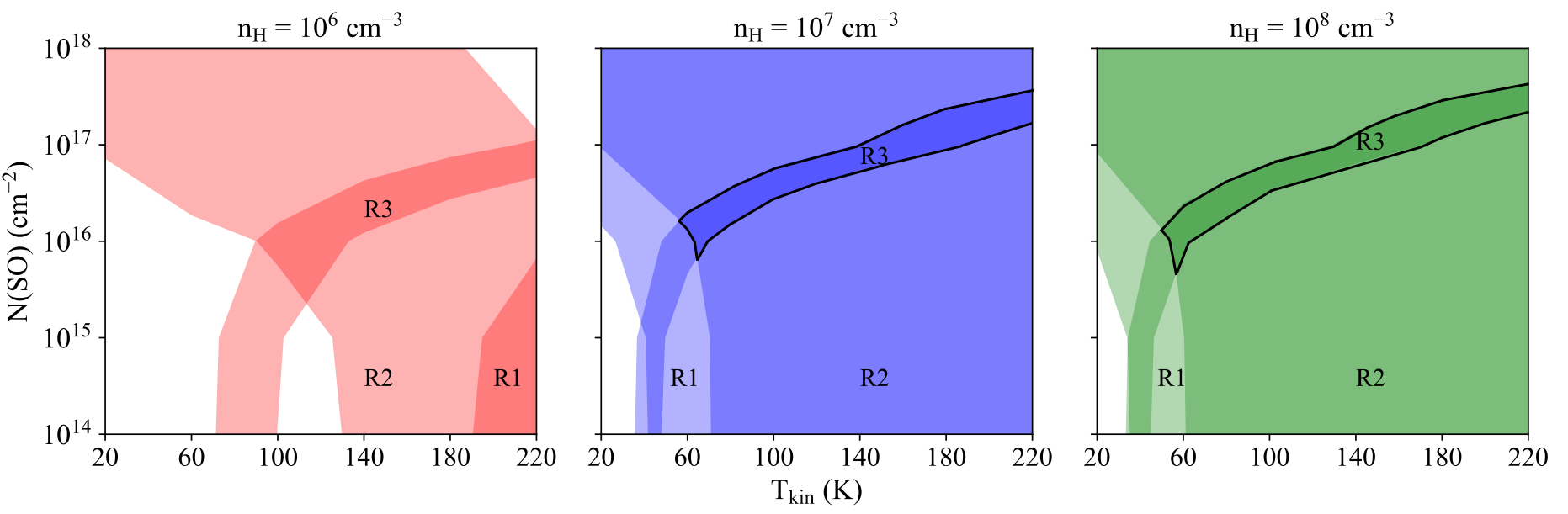}
        \caption{
       	Range of possible values for the SO column density and the kinetic temperature (black contours) from the overlap of the observed intensity ratios in Fig.~\ref{fig:ratios}, for three different values of the H$_{2}$ number density.   
        }
        \label{fig:radex}
\end{figure*}

The intensity ratios between the four brightest SO transitions are presented in Fig.~\ref{fig:ratios} and compared with RADEX models to estimate the kinetic temperature (\textit{T$_\mathrm{kin}$}) and SO column densities. The models are run for three different H$_{2}$ number densities: 10$^{6}$, 10$^{7}$, and 10$^{8}$~cm$^{-3}$. By comparing the observed intensity ratios with the RADEX models, the possible values are presented in Fig.~\ref{fig:radex}. The \textit{n$_\mathrm{H}$}~=~10$^{6}$~cm$^{-3}$ scenario is discarded, given that there is no range of temperatures and densities that contain the three observed line ratios. For \textit{n$_\mathrm{H}$}~=~10$^{7}$~cm$^{-3}$, and \textit{n$_\mathrm{H}$}~=~10$^{8}$~cm$^{-3}$, kinetic temperatures (\textit{T$_\mathrm{kin}$}) between 50 and 220~K are possible, while \textit{N$_\mathrm{SO}$} ranges between 0.06 and 7.0~$\times$~10$^{17}$~cm$^{-2}$ (see Fig.~\ref{fig:radex}). We note that temperatures above 220~K are not shown in Fig.~\ref{fig:radex}, but this upper limit was chosen given that gas-phase SO$_{2}$ formation is more efficient at temperatures below 200~K (see Sect.~5.1). At \textit{T$_\mathrm{kin}$}~=~300~K, \textit{n$_\mathrm{H}$} increases slightly, up to 8.0~$\times$~10$^{17}$~cm$^{-2}$.

\end{appendix}
\end{document}